\documentclass[superscriptaddress,aps, prb, twocolumn,showpacs,showkeys,notitlepage,amsmath, amssymb]{revtex4-2}

\usepackage[utf8]{inputenc}
\usepackage[varg]{txfonts}
\usepackage{bm}
\usepackage{soul}
\usepackage{pbox} % Single load
\usepackage[svgnames]{xcolor} % Consolidated options
\usepackage{lipsum}
\usepackage{lineno}
\usepackage{cancel}
\usepackage{braket}
\usepackage{xspace}
\usepackage{physics}
\usepackage{siunitx}
\usepackage{comment}
\usepackage{upgreek}
\usepackage{colortbl}
\usepackage{graphicx}
\usepackage{rotating}
\usepackage{nicefrac}
\usepackage{multirow}
\usepackage{orcidlink}
\usepackage{longtable}
\usepackage{threeparttable}
\usepackage[normalem]{ulem}
\usepackage[caption=false]{subfig}
\usepackage[referable]{threeparttablex}
\usepackage{booktabs,microtype,afterpage}
\usepackage{hyperref}
\hypersetup{
    colorlinks = true,
    linkcolor = Blue,
    citecolor = Blue,
    urlcolor  = Blue
}

\def \muon {$\mu$SR}

\def \SZRO {Sr$_2$ZnReO$_6$}

\def \abbrDP {DP}

\def \abbrref {Ref.}

\def \abbrfig {Fig.}
\def \abbreqn {Eq.}

\graphicspath{{figures/}} % Location of the graphics files
\DeclareGraphicsExtensions{.pdf,.PDF,png,.PNG,.eps,.EPS}
\begin{document}

\title{Magnetic behavior of the~\texorpdfstring{$5d^1$}{} Re-based double perovskite~\texorpdfstring{\SZRO}{}
}

\author{Muhammad Maikudi Isah\orcidlink{0000-0002-6615-2977}}
%\author{\hspace{1mm}Muhammad Maikudi Isah\orcidlink{0000-0002-6615-2977}}
%\email{muhammadmaikudi.isah@unibo.it}
\affiliation{
Dipartimento di Fisica e Astronomia ``A. Righi'', Universit\'a di Bologna, I-40127 Bologna, Italy}

\author{Biswajit Dalal\orcidlink{0000-0001-8607-485X}}
\affiliation{Research Center for Materials Nanoarchitectonics (MANA), National Institute for Materials Science (NIMS), Namiki 1-1, Tsukuba, Ibaraki 305-0044, Japan}
\affiliation{Department of Physics, Achhruram Memorial College, Jhalda, Purulia, West Bengal, 723202, India}

\author{Xun Kang}
%\author{\hspace{1mm}Xun Kang}
%\email{kang.xun@nims.go.jp}
\affiliation{Research Center for Materials Nanoarchitectonics (MANA), National Institute for Materials Science (NIMS), Namiki 1-1, Tsukuba, Ibaraki 305-0044, Japan}
 
\author{Dario Fiore Mosca\orcidlink{0000-0003-2496-0455}}
\affiliation{University of Vienna, Faculty of Physics, Center for Computational
Materials Science, Kolingasse 14-16, 1090, Vienna, Austria}
%\affiliation{Coll\`ege de France, Universit\'e PSL, 11 place Marcelin Berthelot, 75005 Paris, France} 

\author{\hspace{1mm}Ifeanyi John Onuorah\orcidlink{0000-0001-6358-303}}
\affiliation{
Dipartimento di Scienze Matematiche, Fisiche e Informatiche, Universit\'a di Parma, I-43124 Parma, Italy}

%\author{Valerio Scagnoli\orcidlink{0000-0002-8116-8870}}
\author{Valerio Scagnoli}
\affiliation{Laboratory for Mesoscopic Systems, Department of Materials, ETH Zürich, Zürich,
Switzerland}
\affiliation{PSI Center for Neutron and Muon Sciences, 5232 Villigen PSI, Switzerland}

\author{\hspace{1mm}Pietro Bonf\`a\orcidlink{0000-0001-6358-3037}}
%\affiliation{Dipartimento di Scienze Matematiche, Fisiche e Informatiche, Universit\'a di Parma, I-43124 Parma, Italy} % only one (OLD)
\affiliation{Dipartimento di Fisica, Informatica e Matematica, Universit\'a di Modena e
Reggio Emilia, Via Campi 213/a, I-41125 Modena, Italy}
\affiliation{CNR-NANO S3--Istituto Nanoscienze, I-41125 Modena, Italy}
%\affiliation{Nanoscience Institute, National Research Council (CNR-NANO), 41125 Modena, Italy}

\author{\hspace{1mm}Roberto De Renzi\orcidlink{0000-0002-5015-0061}} 
\affiliation{
Dipartimento di Scienze Matematiche, Fisiche e Informatiche, Universit\'a di Parma, I-43124 Parma, Italy}

\author{Alexei A. Belik\orcidlink{0000-0001-9031-2355}}
\affiliation{Research Center for Materials Nanoarchitectonics (MANA), National Institute for Materials Science (NIMS), Namiki 1-1, Tsukuba, Ibaraki 305-0044, Japan}

\author{Cesare Franchini}
\affiliation{University of Vienna, Faculty of Physics, Center for Computational
Materials Science, Kolingasse 14-16, 1090, Vienna, Austria}
\affiliation{
Dipartimento di Fisica e Astronomia ``A. Righi'', Universit\'a di Bologna, I-40127 Bologna, Italy}

\author{Kazunari Yamaura\orcidlink{0000-0003-0390-8244}}
\email{yamaura.kazunari@nims.go.jp}
\affiliation{Research Center for Materials Nanoarchitectonics (MANA), National Institute for Materials Science (NIMS), Namiki 1-1, Tsukuba, Ibaraki 305-0044, Japan}
\affiliation{Graduate School of Chemical Sciences and Engineering, Hokkaido University,
North 10 West 8, Kita-ku, Sapporo, Hokkaido 060-0810, Japan}

\author{\hspace{1mm}Samuele Sanna\orcidlink{0000-0002-4077-5076}}
\email{s.sanna@unibo.it}
\affiliation{
Dipartimento di Fisica e Astronomia ``A. Righi'', Universit\'a di Bologna, I-40127 Bologna, Italy}

\date{\today}

\begin{abstract}
The subtle interplay between spin-orbit coupling, exchange interactions, and cation ordering can lead to exotic magnetic states in transition-metal ions. We report a comprehensive study of the Re-based (5$d^1$) ordered double perovskite oxide~\SZRO{} combining synchrotron x-ray diffraction (XRD), magnetic susceptibility, muon spin relaxation (\muon{}) measurements, and density functional theory (DFT) calculations. XRD reveals that~\SZRO{} crystallizes in the monoclinic structure (space group $P2_1/n$) at low temperature. Magnetic susceptibility data indicate a transition below $\sim \!13$ K, with $M$--$H$ loops showing ferromagnetic-like hysteresis and an unusually high coercive field of 23 kOe at 2 K. Zero-field \muon{} measurements detect static and spatially disordered internal fields below $T_M \simeq $ 12 K, consistent with a canted antiferromagnetic ground state determined by detailed DFT and force-theorem in Hubbard-I calculations. The reduced high-temperature effective moment ($\sim \!0.76 \ \mu_B$) and very small static moment ($\lesssim \!0.2 \ \mu_B$) derived from~\muon{} analysis and local-field simulations indicate a decisive role of spin-orbit coupling. Through a combined experimental and computational approach we unambiguously determine the canted antiferromagnetic order in~\SZRO{}, showing that a very small ordered moment coexists with an exceptionally large coercivity. These results underscore the crucial role of spin-orbit coupling and orbital ordering, providing new insights into magnetism in 5$d^1$ double perovskites.
\end{abstract}

\pacs{}% PACS, 

\maketitle

\section{\label{sec:intro} Introduction}
The magnetic and orbital ordering in strongly correlated materials plays a crucial role in modern condensed matter physics. In particular, transition-metal ions give rise to a wealth of novel physics properties through the complex interplay of charge, orbital, and spin degrees of freedom~\cite{Kugel_Khomskii_1982}. Among the most extensively studied systems are the $B$-site ordered double perovskite (\abbrDP{}) oxides, with general formula $A_2BB^{\prime}O_6$, where $A$ is an alkaline-earth or rare-earth cation and $B/B^{\prime}$ are transition-metal ions in different oxidation states. These materials have garnered significant interest owing to the influence of spin-orbit coupling (SOC), electronic correlations and crystal field interactions on the electronic and magnetic properties that drive the emergence of exotic quantum phases~\cite{PhysRevB.82.174440,Witczak2014, Pourovskii2025} such as the Mott insulators~\cite{PhysRevLett.99.016404,PhysRevB.100.245141}, Weyl semimetals~\cite{PhysRevB.83.205101, PhysRevLett.107.127205, PhysRevB.85.045124},  half metallicity~\cite{PhysRevLett.98.017204, PhysRevB.59.11159, PhysRevLett.108.177202, PhysRevB.64.125126} and quantum spin liquids~\cite{PhysRevB.96.125109}.

Generally, the $B$-site ordered~\abbrDP{}s have been reported to crystallize in the cubic, monoclinic and tetragonal structures and consist of ordered geometry of corner-shared $BO_6$ and $B^{\prime}O_6$ octahedra network, alternatively forming two interpenetrating face-centered (fcc) lattices, and the $A$ site positioned at the voids between the octahedra~\cite{PhysRevB.69.184412}. The combination of the $B$ and $B'$ sites and the hosting order of the magnetic cations at these sites are crucial determinants of the electronic and magnetic properties, and the nature of the exchange interactions in these compounds.  When $B$ and $B^{\prime}$ sites host magnetic ions, the properties are driven by the  $B\!-\!O\!-\!B^{\prime}$ mediated super-exchange interaction. However, when magnetic ion resides only on  the $B^{\prime}$ site, these compounds often become Mott insulators, since the large separation between neighboring $B^{\prime}$ sites reduces the bandwidth and makes the on-site Coulomb repulsion dominant~\cite{PhysRevB.104.024437}. In such a case, the magnetic interactions are defined by edge-shared network of tetrahedra in a fcc lattice and often exhibit geometric frustration in the presence of antiferromagnetic exchange couplings~\cite{PhysRevB.92.020417,PhysRevB.65.144413,Jana2019}. Further, theoretical studies on analogous~\abbrDP{} compounds subjected to strong SOC with only $B^{\prime}$ magnetic site hosting either $4d^n$ or $5d^n$ electronic state predict the possibility to realize a number of magnetic states including magnetocrystalline anisotropic aligned ferromagnetic (FM) and antiferromagnetic (AFM) states, spin nematic phases, multipolar orders~\cite{PhysRevB.82.174440,Witczak2014,PhysRevB.84.094420} and also canted spin states that are stabilized by Jahn-Teller distortions, owing to the interplay of Heisenberg, non-Heisenberg interactions, and the quadrupolar couplings~\cite{PhysRevB.103.104401}.

In this work, we focus on the Re-based~\abbrDP{}~\SZRO{}, where Sr$^{2+}$, non-magnetic Zn$^{2+}$, and magnetic Re$^{6+}$ (5$d^1$) ions occupy the $A$, $B$ and $B^{\prime}$ sites respectively. \SZRO{} belongs to a class of materials where experimental measurements revealed a wide variety of structural, magnetic and electronic properties. As mentioned in~\abbrref{}~\cite{doi:10.1021acs.inorgchem.6b01933}, these class of materials have no direct correlation between the magnetic ground state and crystal symmetry. For example, Ba$_2$ZnReO$_6$ exhibits canted ferromagnetic order~\cite{Barbosa2022}, Ba$_2$MgReO$_6$ hosts multipolar order~\cite{PhysRevResearch.2.022063}, and Ba$_2$YReO$_6$ displays a spin disordered ground state~\cite{Thompson_2014}, all within the cubic $Fm\bar{3}m$ structure. The diverse behaviors observed in cubic compounds naturally motivate investigations of tetragonally elongated systems, which have drawn even greater interest due to their unusual magnetic properties.
Initially, the isoelectronic compounds Sr$_2$CaReO$_6$ and Sr$_2$MgReO$_6$ were suggested to host spin-glass state below  $\approx\!14$ K and $\approx\!50$ K, respectively~\cite{PhysRevB.65.144413,PhysRevB.68.134410}, but recently resonant x-ray diffraction experiments on a high-quality Sr$_2$MgReO$_6$ sample, have proposed a layered antiferromagnetic order at temperatures below $\approx\!55$ K with a propagation vector $\bm{k} = (0, 0, 1)$~\cite{PhysRevB.101.220412}. Early theoretical studies predicted that these structures could exhibit antiferromagnetic ordering of magnetic octupoles~\cite{PhysRevB.82.174440}, a phenomenon that has evaded direct experimental confirmation. More recently, theory has pointed to Sr$_2$MgReO$_6$ as a specific candidate for realizing such ordering~\cite{Fioremosca2025}. Much less attention has been devoted to understanding the magnetic behavior of~\SZRO{}, including the role of SOC in stabilizing its ground state. Only the existence of contrasting antiferromagnetic features~\cite{PhysRevB.69.184412} and weak ferromagnetic transition~\cite{Retuerto:2008} were detected at low temperature from magnetization measurements, and no magnetic order was detected by neutron diffraction measurements~\cite{Retuerto:2008}. The unusually small magnetic moment is probably a consequence of strong SOC in 5$d^1$~\abbrDP{}s, combined with effects such as covalency, orbital ordering, and Jahn-Teller distortions, all of which contribute to partial cancellation between spin and orbital magnetism~\cite{PhysRevB.104.024437}.

In this paper, we take advantage of high-quality crystalline samples to revise the crystallographic structure of~\SZRO{} using synchrotron x-ray diffraction. We then investigate the magnetization dynamics and magnetic structure through bulk magnetization measurements and muon spin rotation and relaxation (\muon{}), the latter being a highly sensitive local probe of magnetism on the atomic scale and ideally suited to detect the very small magnetic moments predicted in this compound. Finally, we complement our experimental findings with DFT+$\mu$ and force-theorem in Hubbard-I calculations to identify and validate the nature of the magnetic order.

The remainder of this paper is organized as follows; In Sec.~\ref{sec:methods} we present both the experimental and computational methods  that have been utilized in this work followed by the description of the crystal structure obtained from synchrotron XRD in Sec. \ref{sec:crystal}. In Sec.~\ref{sec:magnetic}, the magnetization measurements data are presented while the $\mu$SR measurements together with the DFT calculations of the muon sites are presented in Sec.~\ref{sec:musr} and Sec.~\ref{sec:muon_dipolar}, respectively. The identification of the magnetic structure based on force-theorem in Hubbard-I approach and the validation of the structure by the muon local field simulation is presented in Sec.~\ref{sec:dipolar}. The summary is presented in Sec.~\ref{sec:conclu}. More details on data analysis are presented in the Supplemental Material (SM)~\footnote{See Supplemental Material at [URL will be inserted by publisher], which includes Refs.~\cite{Bonfa2018_muesr, Pourovskii2016, magint, Fioremosca2025, Fioremosca2024b, PhysRevB.103.104401}, for more additional data and details}.

\section{\label{sec:methods} Experimental and Computational Methods}
Polycrystalline~\SZRO{} was synthesized via solid-state reaction using SrO (prepared from 99.9\% pure SrCO$_3$, Wako Pure Chemical Industries, by heating at 1300 $^{\circ}$C in oxygen), ZnO (99.9\%, Wako Pure Chemical Industries), and ReO$_3$ (synthesized in the laboratory from 99.99\% pure Re, Rare Metallic Co. Ltd.). The powders were mixed in an agate mortar inside an argon-filled glovebox to ensure precise stoichiometry. The mixture was sealed in a platinum capsule and subjected to isotropic compression in a multi-anvil press (CTF-MA1500P, C\&T Factory, Tokyo) at 6 GPa and 1100 $^{\circ}$C for 1 h, with an 11-min ramp to reach the target temperature. After heating, the capsule was rapidly quenched to below 100 $^{\circ}$C in 1 min, followed by slow pressure release over several hours. The resulting product was a dense, polycrystalline black pellet. A portion was ground for phase identification using a MiniFlex600 x-ray diffractometer (Rigaku, Tokyo) with Cu-$K{\alpha}$ radiation, and further analyzed by synchrotron XRD with a large Debye-Scherrer camera at BL02B2, SPring-8, Japan~\cite{NISHIBORI20011045, 10.1063/1.4999454}. The x-ray wavelength was $\lambda = 0.65298$~\AA, calibrated using a CeO$_2$ standard. Patterns were analyzed and visualized using the RIETAN-VENUS software package~\cite{Momma:ko5060, Izumi2007ThreeDimensionalVI}.

The direct current (dc) magnetic susceptibility ($\chi$) of~\SZRO{} was measured using a magnetometer (MPMS, Quantum Design, San Diego, CA). The empty sample holder's magnetization was measured to account for its diamagnetic contribution. Measurements were taken from 2--280 K with a 10 kOe magnetic field under both zero-field-cooled (ZFC) and field-cooled (FC) conditions. Isothermal magnetization loops were recorded at various temperatures with the magnetic field swept between $-70$ kOe and $+70$ kOe under ZFC conditions. Data reproducibility was confirmed by testing multiple samples from different synthesis runs.

The zero-field (ZF)~\muon{} measurements were carried out on the GPS spectrometer at the Paul Scherrer Institut (Switzerland) as a function of temperature, ranging from 1.6 K to about 70 K. In a typical ZF-$\mu$SR experiment a beam of positive muons 100\% spin-polarized along the beam direction is implanted in the sample~\cite{TomLancaster2021}. The positive muons thermalize at interstitial sites where they act as sensitive probes of development of spontaneous magnetic ordering, precessing in the local magnetic field, $B_\mu$, at the Larmor frequency $\omega_\mu=\gamma_\mu B_\mu$ ($\gamma_\mu= 2\pi\times135.5~\text{MHz/T})$. By studying the angular distribution of the positrons emitted during the muon decay process (muon lifetime $\tau_\mu\approx2.2 \ \mu  s$) we measure the time evolution of the muon-spin asymmetry $A(t)= A_0 P_z(t)$, where $A_0$ is the initial muon asymmetry and $P_z(t)$ the time dependent muon polarization. The \muon{} experimental data were analyzed via least-squares optimization using MUSRFIT software~\cite{SUTER201269} and MULAB, a home-built MATLAB suite.

To reliably interpret the~\muon{} results, it is crucial to accurately determine the muon stopping site, which we have done via DFT calculations within the DFT$+\mu$ approach~\cite{Moller_2013, bonfa2016,  blundell2023, onuorah2025}. Non-spin polarized DFT calculations were performed using the plane wave (PW) based code Quantum ESPRESSO~\cite{qe2017}. The Perdew-Burke-Ernzerhof (PBE)~\cite{pbe1996} functional was used to estimate the exchange and correlation term. The muon and the host atoms were modeled with ultrasoft pseudopotentials~\cite{PhysRevB.41.7892, GARRITY2014446}  using 100 Ry and 900 Ry cutoff energy for the wavefunctions and charge density respectively. The muon in the DFT+$\mu$ procedure was treated as a hydrogen impurity in a charged $2\times2\times2$ supercell comprising of 160 host atoms and 1 muon. A $2\times2\times2$ Monkhorst-Pack $k$-point mesh~\cite{kpoint1976} was used for the Brillouin zone integration.  The structural relaxations were carried out until forces and total energy differences were less than 1 mRy/Bohr and 0.1 mRy, respectively. All calculations were performed keeping fixed the experimental lattice parameters (monoclinic phase) reported in Sec.~\ref{sec:crystal} below. To resolve the magnetic structure of~\SZRO{}, additional calculation for local magnetic fields at the muon-stopping sites were performed (see SM~\cite{Note1}).

\begin{figure}[t]
    \includegraphics[width = 0.9\linewidth]{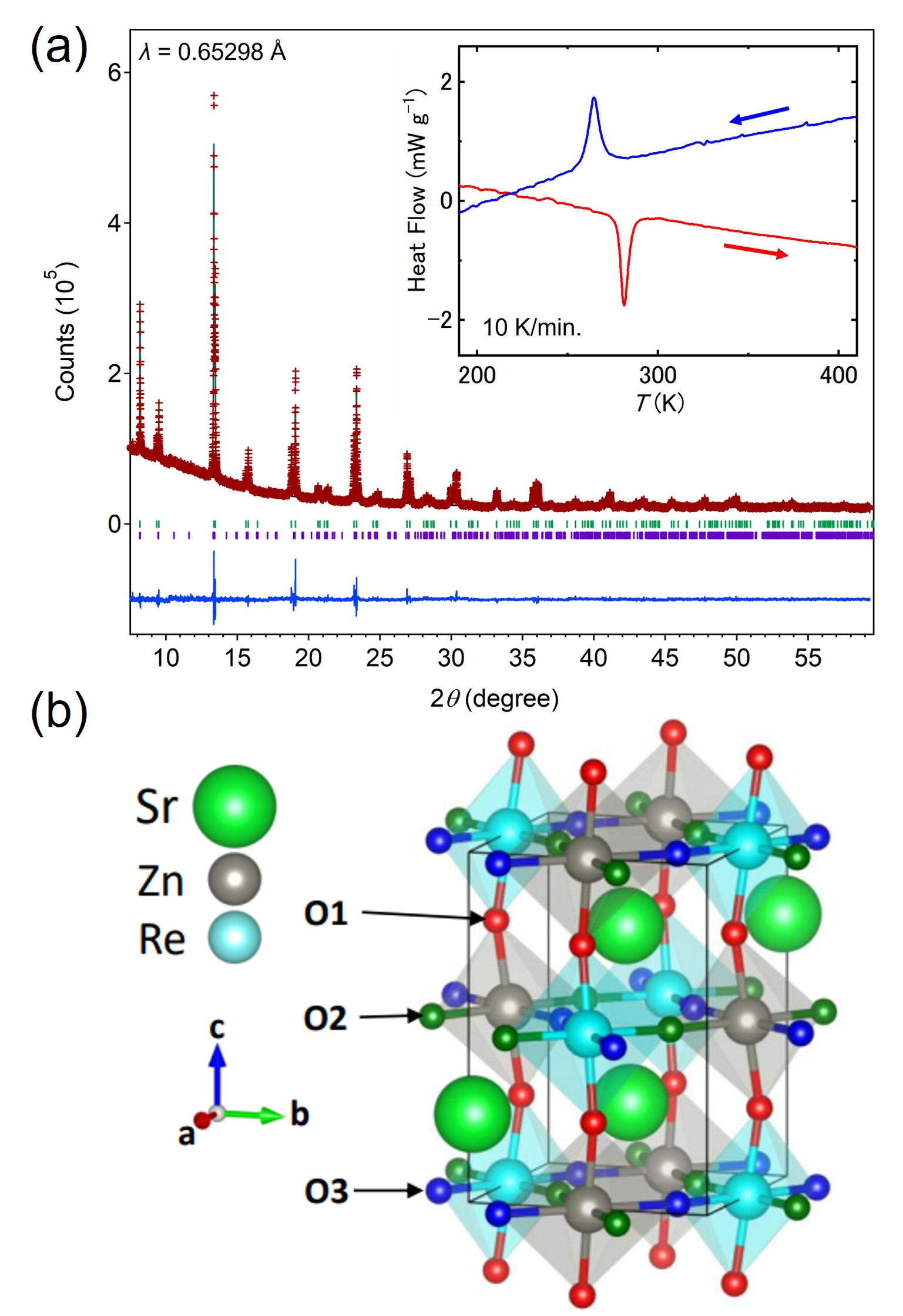}
    \caption{
    (a) Synchrotron XRD pattern of~\SZRO{} at room temperature [298(2) K]. The observed pattern is indicated by red crosses, the calculated pattern by the green solid line, and the difference profile by the blue line. Vertical bars mark the Bragg reflection positions for the tetragonal ($I4/m$; first green ticks) and monoclinic ($P2_1/n$; second violet ticks) models. Inset: DSC heat-flow curves on heating (red) and cooling (blue) measured at $10\ \mathrm{K}\ \mathrm{min}^{-1}$, showing an endothermic/exothermic pair near 300 K with thermal hysteresis.
    (b) Crystal structure of monoclinic~\SZRO{} showing the network of the Zn/ReO$_6$ octahedra. The three inequivalent oxygen atoms are represented with spheres of different colors: O1 (red), O2 (green) and O3 (blue).
    }
    \label{fig:xrd_pattern}
\end{figure}

The magnetic order was determined by calculating the intersite exchange interactions (IEI) for the general low-energy effective many-body Hamiltonian coupling multipolar moments with total angular momentum $J_{\text{eff}} = 3/2$ and subject to a monoclinic crystal field (see SM~\cite{Note1}). The calculation involves several steps. To start, the paramagnetic electronic structure of~\SZRO{} is determined using the charge self-consistent density functional theory with dynamical mean-field theory (DFT+DMFT)~\cite{Georges1996,Anisimov1997_1,Lichtenstein_LDApp,Aichhorn2016} in the quasi-atomic Hubbard-I (HI) approximation~\cite{hubbard_1}. The DFT calculations were performed using the full-potential LAPW method implemented in Wien2k~\cite{Wien2k}, with the SOC effect included via the standard variational treatment. The local density approximation (LDA) was used for the DFT exchange-correlation potential, together with a 500 $k$-point mesh in the full Brillouin zone and a basis cutoff of $R_{\mathrm{mt}}K_{\mathrm{max}} = 7$. The fully localized limit was adopted for the double-counting correction, assuming the nominal 5$d$ shell occupancy of 1. The Wannier orbitals representing the Re $d$ states were constructed from the  $d$ Kohn-Sham bands within the energy window $[-1.36:5.44]$~eV relative to the KS Fermi level and the full $d$-shell parameters were set to $F^0 = U = 3.2$ eV and $J_{H} = 0.5$ eV, consistent with previous studies on $d^1$ and $d^2$~\abbrDP{}s~\cite{PhysRevB.103.104401, Fioremosca2024b, Pourovskii2021}. The calculations were performed with fixed (monoclinic) experimental lattice parameters as reported in Sec.~\ref{sec:crystal}. Our DFT+HI calculation correctly reproduce the effective atomic level scheme, finding a $t_{2g}-e_g$ crystal field splitting of $\sim$4.22 eV and a SO splitting of $\sim$0.5 eV. The $J_{\text{eff}} = 3/2$ ground state multiplet is further split by the monoclinic crystal field in two doublets separated by $\approx$0.17 eV with the corresponding crystal field matrix, reported in the SM, exhibiting a strong mixing of different $m_J$ components due to both the monoclinic distortion and the hybridization with excited $J_{\mathrm{eff}}$=1/2 states. Since theoretical studies predict that multipolar ordering plays a central role in 5$d^1$~\abbrDP{}s, the multipolar IEI was extracted using the force-theorem in Hubbard-I (FT-HI) approach described in Ref.~\cite{Pourovskii2016} via the open-source MagInt code~\cite{magint}. This framework allows the computation of multipolar IEI for general lattice structures with multiple correlated sites and crystal field environments (see SM~\cite{Note1}). All quantities were evaluated in the global reference frame as defined in~\abbrfig{}~\ref{fig:xrd_pattern}(b).

\begin{figure*} [htbp]
    \includegraphics[width = 0.9\linewidth]{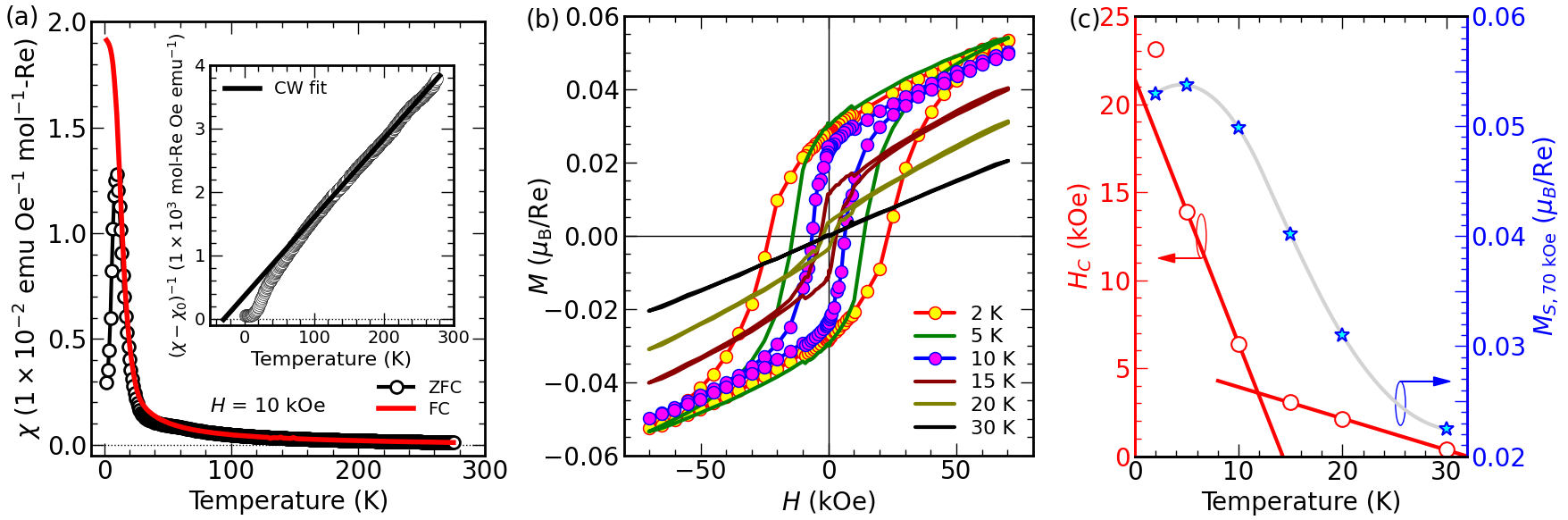}
    \caption{
        (a) Temperature dependence of ZFC (open dot) and FC (solid red line) dc magnetic susceptibility  ($\chi (T)$; main axes) and inverse of ZFC DC magnetic susceptibility with the CW fit (solid line) ($(\chi-\chi_0)^{-1} (T)$; inset axes) in an  applied magnetic field ($H=10$ kOe).
        (b) Isothermal magnetization ($M$) loop as a function of magnetic field ($H$) ranging from -70 to +70 kOe under ZFC mode at six different temperatures.
        (c) Temperature dependence of coercive field ($H_C$; left axes) and saturation magnetization  ($M_S$; right axes) obtained from the ZFC $M(H)$ loops at $H=70$ kOe. 
        The solid lines are guide to the eye unless otherwise stated. 
        %The unit 1 Oe = $10^3/4\pi$ A m$^{-1}$.
        }
        \label{fig:dcmagnetizations}
\end{figure*}

\section{\label{sec:results} Results and Discussion}
\subsection{\label{sec:crystal} Crystal structure}
The synchrotron powder x-ray diffraction patterns showed no detectable impurity peaks, indicating that the samples contained only minimal impurities, if any. Previous studies~\cite{Retuerto:2008, PhysRevB.69.184412} reported a tetragonal phase with the presence of 2.6\% to 15\% of the monoclinic~\SZRO{} phase at room temperature, likely due to variations in synthesis conditions. Therefore, we analyzed the diffractograms of the samples  across a temperature range of 100 K to 300 K, using a model which takes into account both tetragonal (space group $I4/m$; no. 87) and monoclinic (space group $P2_1/n$; no. 14) phases (see ~\abbrfig{}~S1 of the SM~\cite{Note1}). As illustrated in~\abbrfig{} \ref{fig:xrd_pattern}(a), the combined analysis of both models yielded more accurate results than using either model individually. 
Within our experimental uncertainty, no antisite disorder or significant metal/oxygen nonstoichiometry is detected (see Table~S1 note and ~\abbrfig{}~S2 of the SM~\cite{Note1}).
Therefore, in the final analysis, the site occupancy of all atoms was fixed as fully occupied. Figure \ref{fig:xrd_pattern}(b) shows the crystal structure of the monoclinic phase including the network of corner-shared Zn/ReO$_6$ octahedra,  while the quantitative details of the lattice parameters and atomic positions of both phases are tabulated in Table~S1 of the SM~\cite{Note1}.

Having established the structural framework and refined atomic positions, we now turn to the temperature dependence of the crystallographic phases.
At 100 K, the monoclinic phase was dominant, comprising 87\% of the sample. In contrast, at 300 K, the tetragonal phase became dominant, although 24\% of the sample volume is in the monoclinic phase. The phase fraction changed most markedly above 200 K during heating, but the transition remained incomplete even at 300 K. The evolution of the phase fractions shows a steep change near 225 K superimposed on a broad transformation regime extending to $\approx\!300$ K, consistent with a single, first-order-like transition exhibiting extended coexistence rather than multiple transitions. The extended coexistence of tetragonal and monoclinic phases up to 300 K, together with the endothermic/exothermic pair of DSC exhibiting thermal hysteresis [inset to \abbrfig{}~\ref{fig:xrd_pattern}(a)], indicates a structural transition similar to first order. The $\approx 10$\% tetragonal remnant at low $T$ originates from hysteresis of the broad first-order-like transition. Magnetization data--collected on cooling--show no extraneous features attributable to this minority phase. We therefore refer to the transition as first-order-like below. Our results indicate that this transition is coupled with changes in the degree of tilting and buckling of the bonds between the octahedrally coordinated ReO$_6$ and ZnO$_6$ units, in agreement with neutron diffraction studies~\cite{Retuerto:2008}. The thermal evolution of lattice parameters and synchrotron XRD patterns (see \abbrfig{} S3 of the SM~\cite{Note1}) further supports the occurrence of this first-order structural transition over a broad temperature range.

\subsection{\label{sec:magnetic} Magnetic susceptibility and magnetization}
Figure~\ref{fig:dcmagnetizations}(a) presents the temperature dependence of the magnetic susceptibility ($\chi$ vs $T$) in an applied magnetic field of 10 kOe measured on a~\SZRO{} powder sample between 2 K and 280 K. The ZFC curve shows a peak centered at $T_p\sim10$ K; while the FC curve reveals a sharp increase in $\chi$ at $T_M\sim13$ K, implying that the transition at $T_M$ ($T_p$) has a magnetic origin. This can be further corroborated by the dip in the ${d\chi}/{dT}$ versus $T$ curve at around 13 K (not shown here). In addition, the frequency independent nature of the  ac susceptibility peaks at the transition (see~\abbrfig{}~S4 of the SM~\cite{Note1}) indicates the absence of a glassy magnetic state. The inverse magnetic susceptibility, was fitted [inset to~\abbrfig{}~\ref{fig:dcmagnetizations}(a)] in the high-temperature range ($T>100$ K) by assuming a $\chi_0$ temperature-independent contribution due to core diamagnetism and Van Vleck paramagnetism, plus a Curie-Weiss (CW) temperature dependence: $\chi = \chi_0 + {C}/{(T-\Theta_{\mathrm{CW}})}$, where $C$ is the Curie constant and $\Theta_{\mathrm{CW}}$ is the CW temperature. The fit yields $C=0.072(1)$ emu Oe$^{-1}$ K/mol, and $\Theta_{\mathrm{CW}}=-20(1)$ K (see~\abbrfig{}~S5 of the SM~\cite{Note1}).

The paramagnetic effective moment, $\mu_{\mathrm{eff}} = \sqrt{8C}\mu_{\mathrm{B}}=0.758(5)~\mu_{\mathrm{B}}/$Re is smaller than the calculated value, $ \mu_{\mathrm{cal}} = g\sqrt{S(S+1)}\mu_{\mathrm{B}}=1.732~\mu_{\mathrm{B}}$, in the spin only ($S=1/2$) limit and assuming $g=2$. This discrepancy hints at the likely roles played by the neglected orbital moment and the effects of strong SOC in this compound, as obtainable in several $5d^1$~\abbrDP{}s. For instance, in the cubic structured Ba$_2$NaOsO$_6$, the $J_{\mathrm{eff}} = 3/2$ quartet ground state has been established and a corrected $g$-factor arising from the effect of hybridization with the ligands, has been utilized to compute the paramagnetic effective moment~\cite{PhysRevB.95.064416}. The extent of the hybridization is represented by a scale factor $\gamma$, that reduces the effective orbital momentum from the ideal $L_{\mathrm{eff}} = -1$ such that $2S+L_{\mathrm{eff}}\ne0$ and the $g$-factor is $g = 2(1-\gamma)/3$. On this basis, we expect a $\gamma$ value of 0.41 in comparison to $\gamma=0.536$ for Ba$_2$NaOsO$_6$~\cite{PhysRevB.95.064416} and $\gamma=0.49$ for Ba$_2$MgReO$_6$~\cite{doi:10.7566/JPSJ.88.064712} to explain $\mu_{\mathrm{eff}} \sim$0.76 $\mu_{\mathrm{B}}$ for~\SZRO{}.

\begin{figure*}[htbp]%[htbp]
    \includegraphics[width = 1.0\linewidth]{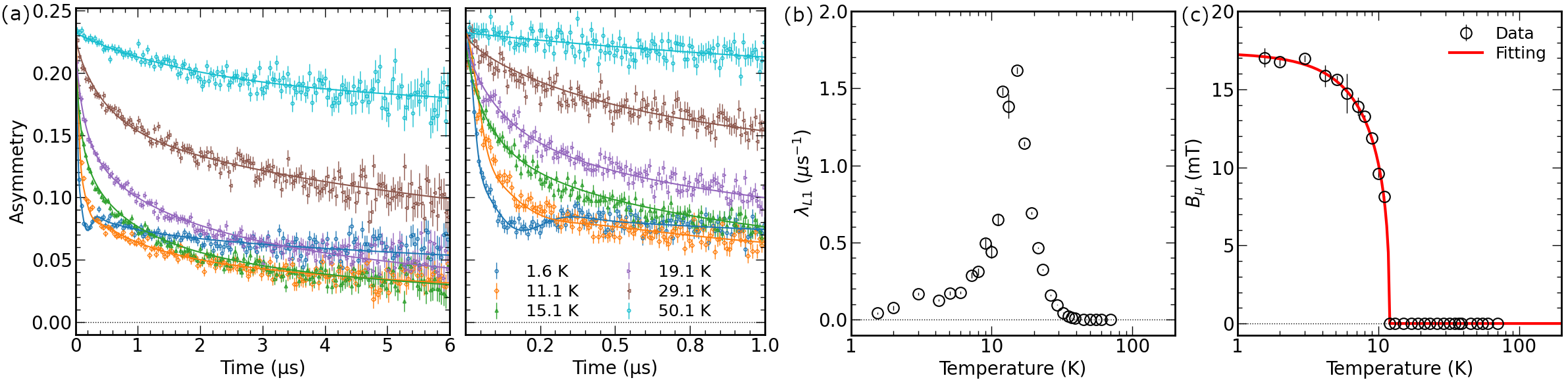}
    \caption{
        (a) ZF-\muon{} asymmetry time spectra at varying temperatures above and below $T_M$. The right panel is a zoom for short acquisition time window. The data are fitted with the function shown in~\abbreqn{}~\eqref{eqn:ZFKT_FIT_EQN}.
        The temperature dependence of the fitted 
        (b) depolarization rate (longitudinal) $\lambda_{L1}$ and
        (c) the internal field $B_{\mu}$ of the oscillatory component, consistent with the magnetic order transition below $T_M$ . The solid red line are fit to the phenomenological function described in the text.
        }
        \label{fig:ZFmuSR}
\end{figure*}

To further elucidate the magnetic behavior of~\SZRO{}, we recorded isothermal field-dependent magnetization $M(H)$ curves under ZFC conditions. Figure~\ref{fig:dcmagnetizations}(b) shows the $M(H)$ curves of~\SZRO{} at selected temperatures. Notably, magnetization curves at low temperatures ($T\lesssim20$ K) display unexpected large hysteresis loops (coercive field $\sim$23 kOe at 2 K) indicating the presence of a hard FM-like character with low magnetic moment, whereas at $T\gtrsim20$ K, it shows a PM behavior. The coercive field ($H_C$) is order of magnitude higher than previously reported~\cite{Retuerto:2008}. 
However, the negative sign of  $\Theta_{\mathrm{CW}}$ suggests an AFM interaction which appears inconsistent with the observed FM behavior. But then, similar negative $\Theta_{\mathrm{CW}}$ has been measured in ferromagnetic Ba$_2$MgReO$_6$, Ba$_2$ZnReO$_6$~\cite{doi:10.1021acs.inorgchem.6b01933} and Ba$_2$NaOsO$_6$~\cite{PhysRevLett.99.016404, PhysRevMaterials.7.084409} compounds, and is attributed to the impact of orbital ordering in the CW behavior of 5$d^1$~\abbrDP{}s, able to stabilize canted ferromagnetic and noncollinear antiferromagnetic orders~\cite{PhysRevB.104.024437}.

In figure~\ref{fig:dcmagnetizations}(c), we plot temperature dependence of coercive field $H_C$ (left axes) and saturation magnetization $M_S$ (right axes). The maximum value obtained for $H_C$ is 23 kOe at 2 K. $H_C$ clearly reveals two slopes as a function of temperature. The slope change occurs at about 13 K, corresponding to a coercivity smaller than the value deduced at 2 K by a factor of $\sim$5. As temperature increases, coercivity slowly decreases reaching a minimum value of 0.36 kOe at 30 K. This may suggest that there is no ordered magnetic phase present for temperatures $T\gtrsim30$ K. The $M(H)$ curves remain unsaturated even at 70 kOe as shown in~\abbrfig{}~\ref{fig:dcmagnetizations}(b), and very small saturated magnetic moment $\sim\!0.05 \ \mu_\mathrm{B}$/Re at 2 K is obtained [\abbrfig{}~\ref{fig:dcmagnetizations}(c)].

To reconcile the unusually large $H_{\mathrm{C}}$ with the tiny moment, we note that strong spin–orbit coupling of Re$^{6+}$ ($5d^{1}$) in a noncubic crystal field can produce substantial magnetocrystalline anisotropy; microstructural pinning from strain/defects may also contribute. To estimate $M_{\mathrm{S}}$, we fitted the high–field region of the $2\ \mathrm{K}$ isotherm with the Law of Approach to Saturation (LAS), $M(H)=M_{\mathrm{S}}\!\left(1-B/H^{2}\right)+\chi H$, obtaining $M_{\mathrm{S}}=0.0566\ \mu_{\mathrm{B}}$/Re (see~\abbrfig{}~S6 of the SM~\cite{Note1}), consistent with the $\sim\! 0.05\ \mu_{\mathrm{B}}$/Re value at $70$ kOe. Because LAS fits are not reliable for extracting precise anisotropy constants in strongly anisotropic magnets, we refrain from quoting $K$ here. Additional diagnostics (e.g., Arrott/Arrott–Noakes analysis and SEM/crystallite-size estimates) will be pursued to separate intrinsic anisotropy from pinning effects. 
The residual slope in $M(H)$ at high field arises from the intrinsic high–field susceptibility of the monoclinic phase rather than a paramagnetic impurity. The minor tetragonal fraction detected by XRD remains paramagnetic and contributes negligibly to the overall magnetization.

\subsection{\label{sec:musr} \texorpdfstring{\muon}{}}
To further probe the magnetic behavior of~\SZRO{}, ZF-\muon{} measurements were performed. The time evolution of muon asymmetry is shown in~\abbrfig{}~\ref{fig:ZFmuSR}(a) for representative temperatures below and above the magnetic transition. A strongly depolarized fraction and spontaneous oscillations associated with the static ordered Re magnetic moment appears below the magnetic transition. The muon asymmetry data, $A(t)$, were modeled by the following function (solid lines in in~\abbrfig{}~\ref{fig:ZFmuSR}) in the whole temperature range ($\chi^2\approx1.1$):

\begin{equation}
\begin{split}
A(t) &=A_T \!\bigg[
       e^{-\frac{\sigma_{T1}^2 t^2}{2}}
       + \eta \cos(\gamma_\mu B_\mu t)
         e^{-\frac{\sigma_{T2}^2 t^2}{2}}
     \bigg] + \\
      & + A_L \!\big[e^{-\big(\lambda_{L1} t\big)^{\beta}} 
       + \eta e^{-\lambda_{L2} t}\big] + A_{\mathrm{bkg}}e^{-\lambda_{\mathrm{bkg}} t}.
\label{eqn:ZFKT_FIT_EQN}
\end{split}
\end{equation}

\noindent
The model takes into account two main muon site fractions (the two terms in the square brackets) with fixed ratio $\eta$ through the all $T$ range plus an additional constant contribution, $A_{bkg}$ with paramagnetic character. At low $T$ the magnetic transition is captured by considering that each of the two fractions develops a transverse component respect to the initial direction of the muon spin $\bm{S}_\mu$ ($A_T$, $\bm{B}_\mu\perp \bm{S}_\mu$) and the longitudinal component ($A_L$, $\bm{B}_\mu\!\!\parallel\!\! \bm{S}_\mu$). The transverse component displays a nonoscillatory Gaussian signal $A_T$ (reflecting overdamped oscillations) plus a single oscillatory damped fraction $\eta A_T$ with Gaussian depolarization function. The longitudinal component requires a fraction $A_L$ with a stretched decay function (with a stretched exponential $\beta$ = 0.5 temperature independent) plus a simple exponential decay $\eta A_L$. The stretched component accounts for a multisite muon population with a distribution of depolarization rates, i.e. of correlation times, that cannot be resolved~\cite{PhysRevB.102.195424}. The best global fit by keeping $A_T+A_L$ constant yields to $\eta=0.24(1)$ consistently for both longitudinal and transverse components through the whole $T$ range. This reflects a muon site population of about 80\% for the site with the longitudinal stretched and highly depolarized transverse contribution, and about 20\% for the other one. Below the magnetic transition the transverse and longitudinal amplitudes (see~\abbrfig{}~S7 of the SM~\cite{Note1}) weights respectively nearly 2/3 and 1/3 of the sum $A_T+A_L$, as expected for a powder sample with a full magnetic volume with static ordered moments. Above the magnetic transition only the $A_L$ component remains recovering the full amplitude. The static character of the magnetic state is also confirmed by longitudinal fields (LF) $\mu$SR measurements (see~\abbrfig{}~S8 of the SM~\cite{Note1}).

%(reported in section SIII of the SM~\cite{Note1}). 

The model also includes a background contribution which turns out to have a constant asymmetry $A_{\mathrm{bkg}}$ and a small constant decay rate of 0.05 $\mu \text{s}^{-1}$ which accounts for the presence of a paramagnetic phase and a possible additional small muon signal coming from the cryostat and sample holder (the latter being typically 1-2\% of the total amplitude on GPS). The global fit returns a $A_{\mathrm{bkg}}$ amplitude of about 12\% of the total signal, a value compatible with the tetragonal paramagnetic fraction measured by XRD.

The estimated temperature dependence of depolarization rate $\lambda_{L1}$ and the internal magnetic field $B_{\mu}$ is presented in~\abbrfig{}~\ref{fig:ZFmuSR}(b-c) (for completeness the behavior of the other free parameters of \abbreqn{}~\ref{eqn:ZFKT_FIT_EQN} are shown in \abbrfig{}~S7 of the SM~\cite{Note1}). As the temperature approaches $T_M$ a sharp increase in $\lambda_{L1}$ is observed, reflecting the critical fluctuations expected at the magnetic transition. The internal magnetic field $B_{\mu}$ [see~\abbrfig{}~\ref{fig:ZFmuSR}(c)] displays a very good agreement with the phenomenological expression:
\begin{equation}
    \label{eqn:power_law}
    B_{\mu}(T) = B_{\mu}(0)\Bigg[1-\Bigg(\frac{T}{T_{M}^{\mu}}\Bigg)^{\alpha}\Bigg]^{\delta}
\end{equation}
\noindent
where $B_{\mu}(0)$ is the zero-temperature internal magnetic field, $\alpha$ is an empirical parameter controlling the saturation behavior, and $\delta$ is the critical exponent-like parameter~\cite{PhysRevB.82.012407, Pelka2013}. The solid red line in~\abbrfig{}~\ref{fig:ZFmuSR}(c) shows that $B_\mu(T)$ follows an order-parameter-like behavior according to~\abbreqn{}~\ref{eqn:power_law} below $T_M$ with $\alpha=1.6(3)$ and $\delta=0.38(4)$. The $\alpha$ value is small (close to the Bloch's $T^{3/2}$ law) and suggests the presence of ferromagnetic-like order~\cite{StephenBlundell2001}, while the $\delta$ value agrees with the expected $\delta \sim 0.367$ for a three-dimensional Heisenberg ferromagnet~\cite{collins1989magnetic}. We also estimate a slightly lower transition temperature $T_{M}^{\mu}=12$~K ($\simeq\!T_M$) if compared to the estimate analysis of our magnetization data, consistent with the zero-field limit probed by \muon{}. The internal magnetic field is $B_{\mu}(0)=17.4(3)$ mT, corresponding to a ground-state frequency of $\nu_{\mu}(0)=2.35(4)$ MHz. This value is consistent with the peak of the field distribution revealed by the Fourier transform of the transverse component signal, discussed in the next section.

\begin{table} %[ht!]
    \caption{
    Candidate muon stopping sites in~\SZRO{} at $4e$ Wyckoff site symmetries. First and second columns contain the site labels. The positions in fractional coordinates are reported in the third column. The fourth column indicates the energy difference with respect to the lowest energy site \textbf{A$_{\mathrm{I}}$}.
    }
    \label{tab:muonsites}
\begin{ruledtabular}
    \begin{tabular}{cccc}
        \begin{tabular}[c]{@{}c@{}} {} \\ {Sites} \end{tabular} 
        &
        \begin{tabular}[c]{@{}c@{}} Label\end{tabular} 
        &
        \begin{tabular}[c]{@{}c@{}} Fractional \\ coord. \end{tabular} 
        &
        \begin{tabular}[c]{@{}c@{}}$\Delta E$ \\ (meV)\end{tabular} 
        \\
        \hline
        \multirow{4}{*}{$\mu$--O1} 
        {} & A$_{\mathrm{I}}$    & (0.068, 0.382, 0.792) & 0 \\
        {} & A$_{\mathrm{II}}$   & (0.409, 0.103, 0.707) & 106 \\
        {} & A$_{\mathrm{III}}$  & (0.853, 0.359, 0.806) & 118 \\
        {} & A$_{\mathrm{IV}}$   & (0.213, 0.302, 0.246) & 149 \\
        \hline
        \multirow{4}{*}{$\mu$--O2} 
        {} & B$_{\mathrm{I}}$    & (0.716, 0.806, 0.592) & 24 \\
        {} & B$_{\mathrm{II}}$   & (0.616, 0.698, 0.501) & 73 \\
        {} & B$_{\mathrm{III}}$  & (0.175, 0.043, 0.519) & 128 \\
        {} & B$_{\mathrm{IV}}$   & (0.801, 0.308, 0.151) & 175 \\
        \hline
        \multirow{4}{*}{$\mu$--O3} 
        {} & C$_{\mathrm{I}}$    & (0.804, 0.103, 0.485) & 32 \\
        {} & C$_{\mathrm{II}}$   & (0.713, 0.209, 0.592) & 48 \\
        {} & C$_{\mathrm{III}}$  & (0.164, 0.301, 0.850) & 144 \\
        {} & C$_{\mathrm{IV}}$   & (0.051, 0.170, 0.993) & 199 \\
    \end{tabular}
\end{ruledtabular}
\end{table}

\begin{figure}[!h]
    \includegraphics[width = 0.9\linewidth]{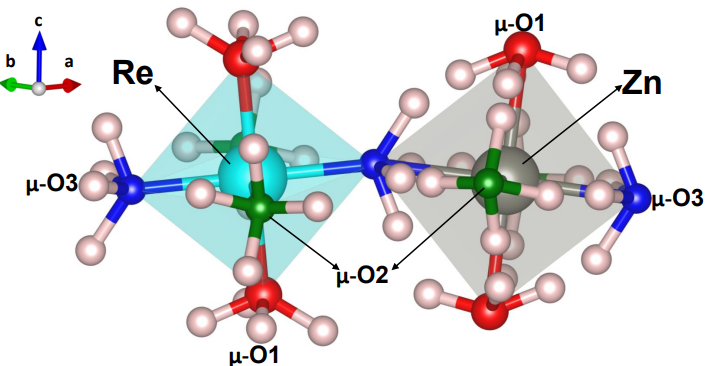}
    \caption{
     Zn/ReO$_6$ (gray/cyan sphere) octahedra showing the muon stopping sites (pink spheres) bonded to the three distinct oxygen sites (red for O1, green for O2 and blue for O3). The $\mu$--$\text{O}_i$ label describes the grouping of inequivalent muons bonded to a distinct O site (see Table \ref{tab:muonsites}). Plots are reproduced using the VESTA program \cite{Momma:ko5060}.
    }\label{fig:visual_muonsites}
\end{figure}

\subsubsection{\label{sec:muon_dipolar} Muon sites}
To further analyze the ZF-\muon{} asymmetry spectra and validate the magnetic structure,  the muon-stopping sites in~\SZRO{} were identified using the standard DFT+$\mu$ protocol. The results reveal twelve candidate crystal symmetry inequivalent muon positions located in the $4e$ Wyckoff position in the unit cell. In each case, the muon site forms a bond with an oxygen atom, with a $\mu$--O bond length of $\approx$1~\AA, as shown in~\abbrfig{}~\ref{fig:visual_muonsites}. The positions (Table~\ref{tab:muonsites}) are grouped into $\mu$--O1,  $\mu$--O2 and $\mu$--O3 sites with respect to the three distinct oxygen. The relative energy differences among all sites are $<0.2$~eV, indicating that all candidate sites are likely to be occupied, even as the muon zero-point motion energy (typically around 0.5~eV~\cite{blundell2023}) is not taken into account. Indeed, the multiple muon sites identified for this compound account for the broad distribution of the local field in the low temperature~\muon{} signal, rather than a single field, consistent with the stretched multisite component of \abbreqn{}~\ref{eqn:ZFKT_FIT_EQN}. The form of this distribution includes local fields that vary for the symmetry inequivalent muons closer to each of the three distinct O$^{2-}$ ion and equivalent muons closer to either the magnetic Re$^{6+}$ ion or the nonmagnetic Zn$^{2+}$ ion in the octahedra network.

\begin{figure}[!h]
    \centering
    \includegraphics[width = 1.0\linewidth]{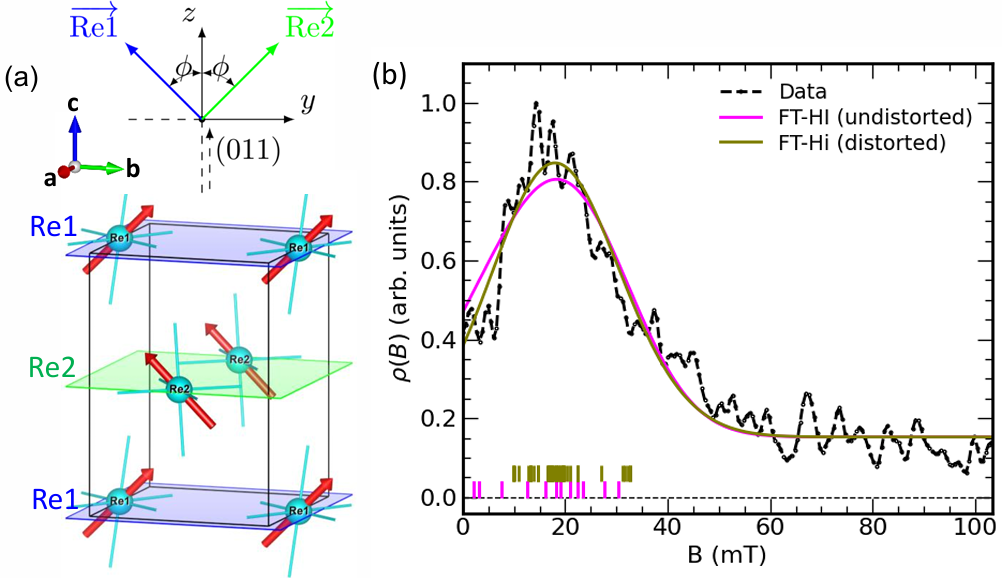} 
    \caption{
    (a) Schematic illustration of the magnetic configuration showing the canting angle $\phi$ with spins aligned along (011) plane and the proposed magnetic structure from the FT-HI calculations with canting angle $\phi= 55^{\circ}$ between the two magnetic sublattices, labeled Re1 (blue plane) and Re2 (green plane).
    (b) Comparison between the experimental ZF-\muon{} field distribution  measured at $T=1.6$ K (black line) and the calculated local field distribution $\rho(B)$. The fits are shown for the proposed \textit{pristine} magnetic structure (``undistorted''; solid magenta line) and for the case including the effects of lattice distortions to the magnetic ions when evaluating the field at the muon site (`distorted''; solid yellow line). The vertical ticks of same color is the calculated local fields at the muon sites using the magnetic moment obtained from the fit.
    }
\label{fig:musr_magnetic_analysis}
\end{figure}

\subsubsection{\label{sec:dipolar}Magnetic structure: FT-HI and muon local field simulations~}
To identify the ground-state magnetic structure, the low-energy effective Hamiltonian was solved in the presence of monoclinic crystal field within the single site mean-field approximation, relying on the IEI calculated from the the FT-HI approach, as described in Sec~\ref{sec:methods}. These calculations were achieved using the ``McPhase'' package~\cite{Rotter2004} together with an in-house module. Due to the quasi-atomic approximation employed, the resulting gyromagnetic factor is $g_J = 0$ because of the exact compensation of  spin and orbital moments, which does not occur in~\SZRO{}. While covalency effects are implicitly included in the DFT+HI calculation via the Wannier orbitals, calculating the orbital magnetic moment from them is a non trivial procedure~\cite{PhysRevB.95.064416} and lies beyond the scope of this work. To compute the magnetic moments, we incorporated the covalency effects through the $\gamma = 1-3g_J/2 = 0.41$ factor described previously in Sec.~\ref{sec:magnetic}. The mean-field result reveal a single second-order phase transition at $\sim$30 K toward a canted antiferromagnetic phase with propagation vector $\bm{k} = (0, 0, 1)$ and net magnetic moment of  $\approx \! 0.04 \ \mu_B$, in very good agreement with the magnetization measurements from~\abbrfig{}~\ref{fig:dcmagnetizations}(b). The overestimation of $T_N$ (in comparison to 12 K of experiment), well known for mean-field approaches is consistent with the approximation used and with previous studies~\cite{Pourovskii2021, Fioremosca2024b}. In~\abbrfig{}~\ref{fig:musr_magnetic_analysis}(a), we show the calculated magnetic order including the schematic description of the observed canting angle $\phi$ between the two Re sublattices. $\phi = 0^{\circ}$ is the FM limit, $\phi = 90^{\circ}$ is the AFM limit, $0^{\circ} < \phi < 45^{\circ}$ angles are closer to FM and thus are defined as the canted FM orders, while those within $45^{\circ} < \phi < 90^{\circ}$ correspond to canted AFM orders. The deduced magnetic configuration consists of spin aligned dominantly in the (011) plane and vanishing contribution along the $x$ direction, with magnetic moments of magnitude $\approx\!0.06 \ \mu_B$ and canting angle $\phi \sim 55^\circ$, suggesting the existence of a canted AFM order. The result was validated against variations of the on-site interaction $U$, confirming the robustness of the solution (see SM~\cite{Note1}).

To elucidate the microscopic origin of the canted AFM phase, we carried out MF calculations in which individual IEI channels were selectively deactivated. The results show that the experimentally observed magnetic structure emerges from the cooperative action of all active time-odd multipolar couplings—dipole–dipole, dipole–octupole, and octupole–octupole (see SM~\cite{Note1}), as previously found in other $5d^1$~\abbrDP{}~\cite{PhysRevB.103.104401}. A quantitative analysis of the role of SOC and crystal-field  effects, left for a future study, is expected to provide additional insights into the origin of this magnetic phase.

To validate the calculated magnetic order and the magnitude of the magnetic moment with the ZF-\muon{} asymmetry spectra, we perform local field simulations at the muon-stopping sites~\cite{PhysRevB.84.144416,PhysRevMaterials.5.044411, PhysRevB.110.064425,advs.202402753}. The local field for a magnetic configuration is calculated using the expression $\mathbf{B} \approx \mathbf{B}_\mathrm{dip} + \mathbf{B}_\mathrm{Lor}$, where the two terms are the dipole field from the magnetic Re$^{6+}$ ions and the Lorentz field respectively~\cite{Bonfa2018_muesr}. The contact hyperfine contribution arising from the overlap between the muon wavefunction and the magnetic Re$^{6+}$ ions is neglected, as vanishing values are expected since the muon positions are relatively far from the magnetic Re$^{6+}$, which also likely host a weak moment. The simulated local field distribution $\rho(B)$ for the obtained canted AFM order is then fitted to the measured field distribution i.e. the Fourier transform power spectrum in the ZF-\muon{} at $T=1.6$ K (see~\abbrfig{}~S7 of the SM~\cite{Note1}). The fitting is discussed in more details in the SM~\cite{Note1}. The result is shown in~\abbrfig{}~\ref{fig:musr_magnetic_analysis}(b), and the simulated field distribution with the proposed canted AFM structure captures the overall features of the experimental data including the peak in the field distribution $\rho(B)$ with $R^2\sim0.91$, signifying the goodness of the fit, validating the deduced magnetic order. A broad distribution of the local field ranging from $\sim$2 mT to $\sim$33 mT is obtained and assigned to  the multiple muon sites. The static magnetic moment deduced from the fit is very small, 0.222(5) $\mu_B$/Re for the pristine \SZRO{} structure (``undistorted'') and slightly reduced to 0.193(3) $\mu_B$/Re when muon‑induced lattice distortions are included (``distorted'') [\abbrfig{}~\ref{fig:musr_magnetic_analysis}(b)] (see~\abbrfig{}~S9 of the SM~\cite{Note1}). Taking into account systematic errors in our analysis and the neglected contact contribution to the local field at the muon sites, a more reliable estimate for the moment at Re is $\lesssim$0.2$\ \mu_B$. This is still larger than the HI estimate, possibly because the hybridization effects are not explicitly taken into account in our calculations, which are known to be strong in this class of materials~\cite{PhysRevLett.133.066501}. As a consequence, it is possible that the effective moment due to hybridization is different and this can affect both  DFT+HI estimates as well as DFT+$\mu$ results through e.g., oxygen polarization.

In all, the obtained canted AFM order is in agreement with the magnetization measurements and reconciles the observed varying FM and AFM signatures in the experimental data. The coexistence of both the large coercive field $\sim$23 kOe at 2 K, indicative of FM-like behavior with low magnetic moment and the negative sign of  $\Theta_{\mathrm{CW}}$ suggesting an AFM interaction is attributed to the impact of orbital ordering in the CW behavior of $5d^1$~\abbrDP{}s~\cite{PhysRevB.104.024437}, able to stabilize the obtained canted AFM order.

\section{\label{sec:conclu}Summary}
In summary, we have investigated the crystal structure and magnetic properties of polycrystalline~\SZRO{}. The crystal structure at room temperature is defined in a tetragonal unit cell (space group $I4/m$) while at low temperature,~\SZRO{} undergoes a reduction in symmetry to monoclinic ($P2_1/n$) phase. The magnetic susceptibility data show a sharp increase indicative of a magnetic phase transition below 13 K which is corroborated by ZF-\muon{} measurements where divergence in the temperature dependence of the relaxation rates has been observed at $T_M\!\simeq\!$ 12 K, below which the existence of static but spatially disordered internal magnetic field, revealed through the presence of multiple muon sites has been measured. $M$--$H$ curves indicate an unusually large hysteresis loop with coercive field of 23 kOe at 2 K indicative of  a FM-like behavior. On the contrary the fit of the magnetic susceptibility above the magnetic transition show a negative Curie-Wiess constant, $\Theta_{\mathrm{CW}}=-20$ K indicating AFM correlations.

A canted AFM order was obtained from DFT and force-theorem in Hubbard-I calculations and validated by the simulation of the muon experimental data. The obtained magnetic configuration consists of spin aligned dominantly in the (011) plane and vanishing contribution along $x$ direction, with magnetic moments of magnitude $\approx$0.06$ \ \mu_B$ and canting angle $\phi \sim 55^{\circ}$. This magnetic order reconciles the hysteretic FM-like behaviour and the AFM character observed with the negative sign of Curie-Wiess temperature.

A small static ordered magnetic moment, of the order of $0.2 \ \mu_{B}$/Re, was obtained from the analysis of the muon data, which neglects hybridization effects and can be considered an upper limit. This is further reduced with respect to the paramagnetic effective moment of $0.76 \ \mu_{B}$ obtained from Curie-Wiess fitting of the magnetic susceptibility measurements, and consistent with reduced moment obtainable in $5d^1$~\abbrDP{}s owing to effect of SOC. To account for the reduced effective moment with respect to the spin only $1.73 \ \mu_{B}$ value, the $J_\mathrm{eff}=3/2$ quartet-like ground state has been assumed together with covalency effects on the cation orbital moment, underlining the impact of SOC in this compound. These results provide evidence for the magnetic ground state of~\SZRO{} and guide to the theoretical description of the strong SOC effects envisaged for this and similar compounds.

\begin{acknowledgments}
The work here presented is partly supported by project ``Spin-charge-lattice coupling in relativistic Mott insulators'' (ID No. 202243JHMW and CUP No. J53D23001350006), funded by European Union--Next Generation EU project--``PNRR--M4C2, investimento 1.1--Fondo PRIN 2022''.This work was also partially supported by a Grant-in-Aid for Scientific Research (Grant No. JP25K01657) from the Japan Society for the Promotion of Science. Synchrotron radiation experiments were conducted at the former NIMS beamline BL15XU of SPring-8 with the permission from the former NIMS synchrotron X-ray station (Proposal No. 2020A4501). We thank Dr. Y. Katsuya and Dr. M. Tanaka for their help at BL15XU of SPring-8. I.J.O. and R.D.R. acknowledge financial support from the PNRR MUR Project No. ECS-00000033-ECOSISTER. This research was partially granted by University of Parma through the action Bando di Ateneo 2023 per la ricerca. This work is based on experiments performed at the Swiss Muon Source S$\mu$S, Paul Scherrer Institute, Villigen, Switzerland. C.F. acknowledges the project ``Superlattices of relativistic oxides'' (ID No. 2022L28H97) funded by European Union--Next Generation EU project--``PNRR--M4C2, investimento 1.1--Fondo PRIN 2022''. M.M.I. acknowledges the Computing resources provided by STFC Scientific Computing Department’s SCARF cluster. D.F.M. thanks the computational facilities of the Austrian Scientific Computing (ASC).
\end{acknowledgments}

\section*{Data Availability}
The data that support the findings of this article are not
publicly available because they contain sensitive personal
information. The data are available from the authors upon
reasonable request.

\bibliography{main}

@article{advs.202402753,
  author       = {Sahoo, Manaswini and Onuorah, Ifeanyi John and Folkers, Laura Christina and Kochetkova, Ekaterina and Chulkov, Evgueni V. and Otrokov, Mikhail M. and Aliev, Ziya S. and Amiraslanov, Imamaddin R. and Wolter, Anja U. B. and Büchner, Bernd and Corredor, Laura Teresa and Wang, Chennan and Salman, Zaher and Isaeva, Anna and De Renzi, Roberto and Allodi, Giuseppe},
  title        = {{Ubiquitous Order-Disorder Transition in the Mn Antisite Sublattice of the (MnBi$_2$Te$_4$)(Bi$_2$Te$_3$)$_n$ Magnetic Topological Insulators}},
  year         = 2024,
  journal      = {Advanced Science},
  volume       = 11,
  number       = 34,
  pages        = 2402753,
  doi          = {https://doi.org/10.1002/advs.202402753},
  url          = {https://advanced.onlinelibrary.wiley.com/doi/abs/10.1002/advs.202402753},
}

@article{PhysRevMaterials.5.044411,
  author       = {Bonf\`a, Pietro and Isah, Muhammad Maikudi and Frandsen, Benjamin A. and Gibson, Ethan J. and Br\"uck, Ekkes and Onuorah, Ifeanyi John and De Renzi, Roberto and Allodi, Giuseppe},
  title        = {{Ab initio modeling and experimental investigation of ${\mathrm{Fe}}_{2}\mathrm{P}$ by DFT and spin spectroscopies}},
  year         = 2021,
  month        = {Apr},
  journal      = {Phys. Rev. Mater.},
  publisher    = {American Physical Society},
  volume       = 5,
  pages        = {044411},
  doi          = {10.1103/PhysRevMaterials.5.044411},
  url          = {https://link.aps.org/doi/10.1103/PhysRevMaterials.5.044411},
  issue        = 4,
  numpages     = 7
}

@article{PhysRevB.100.245141,
  author       = {Cong, R. and Nanguneri, Ravindra and Rubenstein, Brenda and Mitrovi\ifmmode \acute{c}\else \'{c}\fi{}, V. F.},
  title        = {{Evidence from first-principles calculations for orbital ordering in Ba$_2$NaOsO$_6$: A Mott insulator with strong spin-orbit coupling}},
  year         = 2019,
  month        = {Dec},
  journal      = {Phys. Rev. B},
  publisher    = {American Physical Society},
  volume       = 100,
  pages        = 245141,
  doi          = {10.1103/PhysRevB.100.245141},
  url          = {https://link.aps.org/doi/10.1103/PhysRevB.100.245141},
  issue        = 24,
  numpages     = 6
}

@article{Thompson_2014,
  author       = {Thompson, C M and Carlo, J P and Flacau, R and Aharen, T and Leahy, I A and Pollichemi, J R and Munsie, T J S and Medina, T and Luke, G M and Munevar, J and Cheung, S and Goko, T and Uemura, Y J and Greedan, J E},
  title        = {{Long-range magnetic order in the 5d2 double perovskite Ba$_2$CaOsO$_6$: comparison with spin-disordered Ba$_2$YReO$_6$}},
  year         = 2014,
  month        = {jul},
  journal      = {Journal of Physics: Condensed Matter},
  publisher    = {IOP Publishing},
  volume       = 26,
  number       = 30,
  pages        = 306003,
  doi          = {10.1088/0953-8984/26/30/306003},
  url          = {https://dx.doi.org/10.1088/0953-8984/26/30/306003}
}

@article{PhysRevLett.99.016404,
  author       = {Erickson, A. S. and Misra, S. and Miller, G. J. and Gupta, R. R. and Schlesinger, Z. and Harrison, W. A. and Kim, J. M. and Fisher, I. R.},
  title        = {{Ferromagnetism in the Mott Insulator ${\mathrm{Ba}}_{2}{\mathrm{NaOsO}}_{6}$}},
  year         = 2007,
  month        = {Jul},
  journal      = {Phys. Rev. Lett.},
  publisher    = {American Physical Society},
  volume       = 99,
  pages        = {016404},
  doi          = {10.1103/PhysRevLett.99.016404},
  url          = {https://link.aps.org/doi/10.1103/PhysRevLett.99.016404},
  issue        = 1,
  numpages     = 4
}

@article{PhysRevB.92.020417,
  author       = {Cook, A. M. and Matern, S. and Hickey, C. and Aczel, A. A. and Paramekanti, A.},
  title        = {{Spin-orbit coupled ${j}_{\mathrm{eff}}=1/2$ iridium moments on the geometrically frustrated fcc lattice}},
  year         = 2015,
  month        = {Jul},
  journal      = {Phys. Rev. B},
  publisher    = {American Physical Society},
  volume       = 92,
  pages        = {020417},
  doi          = {10.1103/PhysRevB.92.020417},
  url          = {https://link.aps.org/doi/10.1103/PhysRevB.92.020417},
  issue        = 2,
  numpages     = 6
}

@article{Jana2019,
  author       = {Jana, Somnath and Aich, Payel and Kumar, P. Anil and Forslund, O. K. and Nocerino, E. and Pomjakushin, V. and M{\aa}nsson, M. and Sassa, Y. and Svedlindh, Peter and Karis, Olof and Siruguri, Vasudeva and Ray, Sugata},
  title        = {{Revisiting Goodenough-Kanamori rules in a new series of double perovskites LaSr$_{1-x}$Ca$_x$NiReO$_6$}},
  year         = 2019,
  month        = {Dec},
  day          = {04},
  journal      = {Scientific Reports},
  volume       = 9,
  number       = 1,
  pages        = 18296,
  doi          = {10.1038/s41598-019-54427-0},
  issn         = {2045-2322},
  url          = {https://doi.org/10.1038/s41598-019-54427-0}
}

@article{PhysRevB.103.104401,
  author       = {Fiore Mosca, Dario and Pourovskii, Leonid V. and Kim, Beom Hyun and Liu, Peitao and Sanna, Samuele and Boscherini, Federico and Khmelevskyi, Sergii and Franchini, Cesare},
  title        = {{Interplay between multipolar spin interactions, Jahn-Teller effect, and electronic correlation in a ${J}_{\text{eff}}=\frac{3}{2}$ insulator}},
  year         = 2021,
  month        = {Mar},
  journal      = {Phys. Rev. B},
  publisher    = {American Physical Society},
  volume       = 103,
  pages        = 104401,
  doi          = {10.1103/PhysRevB.103.104401},
  url          = {https://link.aps.org/doi/10.1103/PhysRevB.103.104401},
  issue        = 10,
  numpages     = 6
}

@article{onuorah2025,
  author       = {Onuorah, Ifeanyi J. and Bonacci, Miki and Isah, Muhammad M. and Mazzani, Marcello and De Renzi, Roberto and Pizzi, Giovanni and Bonfà, Pietro},
  title        = {{Automated computational workflows for muon spin spectroscopy}},
  year         = 2025,
  journal      = {Digital Discovery},
  publisher    = {RSC},
  volume       = 4,
  pages        = {523--538},
  doi          = {10.1039/D4DD00314D},
  url          = {http://dx.doi.org/10.1039/D4DD00314D},
  issue        = 2
}

@book{StephenBlundell2001,
  author       = {Blundell, Stephen},
  title        = {{Magnetism in Condensed Matter}},
  year         = 2001,
  publisher    = {Oxford University Press},
  address      = {Oxford, United Kingdom}
}

@book{collins1989magnetic,
  author       = {Malcolm F. Collins},
  title        = {{Magnetic Critical Scattering}},
  year         = 1989,
  publisher    = {Oxford University Press},
  address      = {New York},
  pages        = 29
}

@book{TomLancaster2021,
  author       = {Blundell, Stephen J. and De Renzi, Roberto and Lancaster, Tom and Pratt, Francis L.},
  title        = {{Introduction to Muon Spectroscopy}},
  year         = 2022,
  publisher    = {Oxford University Press},
  address      = {Oxford, United Kingdom},
  doi          = {10.1093/oso/9780198858959.001.0001},
  isbn         = 9780198858959,
  url          = {https://doi.org/10.1093/oso/9780198858959.001.0001}
}

@article{Bonfa2018_muesr,
  author       = {Bonf{\`a}, Pietro and Onuorah, Ifeanyi John and De Renzi, Roberto},
  title        = {{Introduction and a Quick Look at MUESR, the Magnetic Structure and mUon Embedding Site Refinement Suite}},
  year         = 2018,
  journal      = {JPS Conf. Proc.},
  volume       = 21,
  pages        = {011052},
  doi          = {10.7566/JPSCP.21.011052},
  url          = {https://journals.jps.jp/doi/10.7566/JPSCP.21.011052}
}

@article{Witczak2014,
  author       = {Witczak-Krempa, William and Chen, Gang and Kim, Yong Baek and Balents, Leon},
  title        = {{Correlated Quantum Phenomena in the Strong Spin-Orbit Regime}},
  year         = 2014,
  journal      = {Annual Review of Condensed Matter Physics},
  publisher    = {Annual Reviews},
  volume       = 5,
  number       = {Volume 5, 2014},
  pages        = {57--82},
  doi          = {https://doi.org/10.1146/annurev-conmatphys-020911-125138},
  issn         = {1947-5462},
  url          = {https://www.annualreviews.org/content/journals/10.1146/annurev-conmatphys-020911-125138},
  type         = {Journal Article}
}

@article{PhysRevResearch.2.022063,
  author       = {Hirai, Daigorou and Sagayama, Hajime and Gao, Shang and Ohsumi, Hiroyuki and Chen, Gang and Arima, Taka-hisa and Hiroi, Zenji},
  title        = {{Detection of multipolar orders in the spin-orbit-coupled $5d$ Mott insulator $\mathrm{B}{\mathrm{a}}_{2}\mathrm{MgRe}{\mathrm{O}}_{6}$}},
  year         = 2020,
  month        = {Jun},
  journal      = {Phys. Rev. Res.},
  publisher    = {American Physical Society},
  volume       = 2,
  pages        = {022063},
  doi          = {10.1103/PhysRevResearch.2.022063},
  url          = {https://link.aps.org/doi/10.1103/PhysRevResearch.2.022063},
  issue        = 2,
  numpages     = 6
}

@article{doi:10.7566/JPSJ.88.064712,
  author       = {Hirai ,Daigorou and Hiroi ,Zenji},
  title        = {{Successive Symmetry Breaking in a $J_{\mathrm{eff}} = 3/2$ Quartet in the Spin–Orbit Coupled Insulator Ba$_2$MgReO$_6$}},
  year         = 2019,
  journal      = {Journal of the Physical Society of Japan},
  volume       = 88,
  number       = 6,
  pages        = {064712}, 
  doi          = {10.7566/JPSJ.88.064712},
  url          = {https://doi.org/10.7566/JPSJ.88.064712}
}

@article{PhysRevB.83.205101,
  author       = {Wan, Xiangang and Turner, Ari M. and Vishwanath, Ashvin and Savrasov, Sergey Y.},
  title        = {{Topological semimetal and Fermi-arc surface states in the electronic structure of pyrochlore iridates}},
  year         = 2011,
  month        = {May},
  journal      = {Phys. Rev. B},
  publisher    = {American Physical Society},
  volume       = 83,
  pages        = 205101,
  doi          = {10.1103/PhysRevB.83.205101},
  url          = {https://link.aps.org/doi/10.1103/PhysRevB.83.205101},
  issue        = 20,
  numpages     = 9
}

@article{PhysRevLett.107.127205,
  author       = {Burkov, A. A. and Balents, Leon},
  title        = {{Weyl Semimetal in a Topological Insulator Multilayer}},
  year         = 2011,
  month        = {Sep},
  journal      = {Phys. Rev. Lett.},
  publisher    = {American Physical Society},
  volume       = 107,
  pages        = 127205,
  doi          = {10.1103/PhysRevLett.107.127205},
  url          = {https://link.aps.org/doi/10.1103/PhysRevLett.107.127205},
  issue        = 12,
  numpages     = 4
}

@article{PhysRevB.85.045124,
  author       = {Witczak-Krempa, William and Kim, Yong Baek},
  title        = {{Topological and magnetic phases of interacting electrons in the pyrochlore iridates}},
  year         = 2012,
  month        = {Jan},
  journal      = {Phys. Rev. B},
  publisher    = {American Physical Society},
  volume       = 85,
  pages        = {045124},
  doi          = {10.1103/PhysRevB.85.045124},
  url          = {https://link.aps.org/doi/10.1103/PhysRevB.85.045124},
  issue        = 4,
  numpages     = 6
}

@article{PhysRevB.82.174440,
  author       = {Chen, Gang and Pereira, Rodrigo and Balents, Leon},
  title        = {{Exotic phases induced by strong spin-orbit coupling in ordered double perovskites}},
  year         = 2010,
  month        = {Nov},
  journal      = {Phys. Rev. B},
  publisher    = {American Physical Society},
  volume       = 82,
  pages        = 174440,
  doi          = {10.1103/PhysRevB.82.174440},
  url          = {https://link.aps.org/doi/10.1103/PhysRevB.82.174440},
  issue        = 17,
  numpages     = 25
}

@article{PhysRevB.84.094420,
  author       = {Chen, Gang and Balents, Leon},
  title        = {{Spin-orbit coupling in ${d}^{2}$ ordered double perovskites}},
  year         = 2011,
  month        = {Sep},
  journal      = {Phys. Rev. B},
  publisher    = {American Physical Society},
  volume       = 84,
  pages        = {094420},
  doi          = {10.1103/PhysRevB.84.094420},
  url          = {https://link.aps.org/doi/10.1103/PhysRevB.84.094420},
  issue        = 9,
  numpages     = 13
}

@article{PhysRevB.96.125109,
  author       = {Natori, W. M. H. and Daghofer, M. and Pereira, R. G.},
  title        = {{Dynamics of a $j=\frac{3}{2}$ quantum spin liquid}},
  year         = 2017,
  month        = {Sep},
  journal      = {Phys. Rev. B},
  publisher    = {American Physical Society},
  volume       = 96,
  pages        = 125109,
  doi          = {10.1103/PhysRevB.96.125109},
  url          = {https://link.aps.org/doi/10.1103/PhysRevB.96.125109},
  issue        = 12,
  numpages     = 18
}

@article{PhysRevB.101.220412,
  author       = {Gao, Shang and Hirai, Daigorou and Sagayama, Hajime and Ohsumi, Hiroyuki and Hiroi, Zenji and Arima, Taka-hisa},
  title        = {{Antiferromagnetic long-range order in the $5{d}^{1}$ double-perovskite ${\mathrm{Sr}}_{2}{\mathrm{MgReO}}_{6}$}},
  year         = 2020,
  month        = {Jun},
  journal      = {Phys. Rev. B},
  publisher    = {American Physical Society},
  volume       = 101,
  pages        = 220412,
  doi          = {10.1103/PhysRevB.101.220412},
  url          = {https://link.aps.org/doi/10.1103/PhysRevB.101.220412},
  issue        = 22,
  numpages     = 6
}

@article{PhysRevB.68.134410,
  author       = {Wiebe, C. R. and Greedan, J. E. and Kyriakou, P. P. and Luke, G. M. and Gardner, J. S. and Fukaya, A. and Gat-Malureanu, I. M. and Russo, P. L. and Savici, A. T. and Uemura, Y. J.},
  title        = {{Frustration-driven spin freezing in the $S=\frac{1}{2}$ fcc perovskite ${\mathrm{Sr}}_{2}{\mathrm{MgReO}}_{6}$}},
  year         = 2003,
  month        = {Oct},
  journal      = {Phys. Rev. B},
  publisher    = {American Physical Society},
  volume       = 68,
  pages        = 134410,
  doi          = {10.1103/PhysRevB.68.134410},
  url          = {https://link.aps.org/doi/10.1103/PhysRevB.68.134410},
  issue        = 13,
  numpages     = 10
}

@article{doi:10.1021acs.inorgchem.6b01933,
  author       = {Marjerrison, Casey A. and Thompson, Corey M. and Sala, Gabrielle and Maharaj, Dalini D. and Kermarrec, Edwin and Cai, Yipeng and Hallas, Alannah M. and Wilson, Murray N. and Munsie, Timothy J. S. and Granroth, Garrett E. and Flacau, Roxana and Greedan, John E. and Gaulin, Bruce D. and Luke, Graeme M.},
  title        = {{Cubic Re$^{6+}$ ($5d^1$) Double Perovskites, Ba$_2$MgReO$_6$, Ba$_2$ZnReO$_6$, and Ba$_2$Y$_{2/3}$ReO$_6$: Magnetism, Heat Capacity, $\mu$SR, and Neutron Scattering Studies and Comparison with Theory}},
  year         = 2016,
  journal      = {Inorganic Chemistry},
  volume       = 55,
  number       = 20,
  pages        = {10701--10713},
  doi          = {10.1021/acs.inorgchem.6b01933},
  url          = {https://doi.org/10.1021/acs.inorgchem.6b01933},
}

@article{Retuerto:2008,
  author       = {Retuerto, María and Martínez-Lope, María Jesús and García-Hernández, Mar and Fernández-Díaz, María Teresa and Alonso, José Antonio},
  title        = {{Crystal and Magnetic Structure of Sr$_2$MReO6 (M = Ni, Co, Zn) Double Perovskites: A Neutron Diffraction Study}},
  year         = 2008,
  journal      = {European Journal of Inorganic Chemistry},
  volume       = 2008,
  number       = 4,
  pages        = {588--595},
  doi          = {https://doi.org/10.1002/ejic.200700753},
  url          = {https://chemistry-europe.onlinelibrary.wiley.com/doi/abs/10.1002/ejic.200700753}
}

@article{PhysRevB.69.184412,
  author       = {Kato, H. and Okuda, T. and Okimoto, Y. and Tomioka, Y. and Oikawa, K. and Kamiyama, T. and Tokura, Y.},
  title        = {{Structural and electronic properties of the ordered double perovskites ${A}_{2}M{\mathrm{ReO}}_{6}$ $(A$=Sr,Ca; $M$=Mg,Sc,Cr,Mn,Fe,Co,Ni,Zn)}},
  year         = 2004,
  month        = {May},
  journal      = {Phys. Rev. B},
  publisher    = {American Physical Society},
  volume       = 69,
  pages        = 184412,
  doi          = {10.1103/PhysRevB.69.184412},
  url          = {https://link.aps.org/doi/10.1103/PhysRevB.69.184412},
  issue        = 18,
  numpages     = 8
}

@article{SUTER201269,
  author       = {A. Suter and B.M. Wojek},
  title        = {{Musrfit: A Free Platform-Independent Framework for $\mu$SR Data Analysis}},
  year         = 2012,
  journal      = {Physics Procedia},
  volume       = 30,
  pages        = {69--73},
  doi          = {https://doi.org/10.1016/j.phpro.2012.04.042},
  issn         = {1875-3892},
  url          = {https://www.sciencedirect.com/science/article/pii/S187538921201228X},
  note         = {12th International Conference on Muon Spin Rotation, Relaxation and Resonance ($\mu$SR2011)}
}

@article{blundell2023,
  author       = {Blundell, S. J. and Lancaster, T.},
  title        = {{DFT + $\mu$: Density functional theory for muon site determination}},
  year         = 2023,
  month        = {06},
  journal      = {Applied Physics Reviews},
  volume       = 10,
  number       = 2,
  pages        = {021316},
  doi          = {10.1063/5.0149080},
  issn         = {1931-9401},
  url          = {https://doi.org/10.1063/5.0149080},
}

@article{Moller_2013,
  author       = {J S Möller and P Bonfà and D Ceresoli and F Bernardini and S J Blundell and T Lancaster and R De Renzi and N Marzari and I Watanabe and S Sulaiman and M I Mohamed-Ibrahim},
  title        = {{Playing quantum hide-and-seek with the muon: localizing muon stopping sites}},
  year         = 2013,
  month        = {dec},
  journal      = {Physica Scripta},
  publisher    = {IOP Publishing},
  volume       = 88,
  number       = 6,
  pages        = {068510},
  url          = {https://dx.doi.org/10.1088/0031-8949/88/06/068510}
}

@article{bonfa2016,
  author       = {Bonf\`{a} ,Pietro and De Renzi ,Roberto},
  title        = {{Toward the Computational Prediction of Muon Sites and Interaction Parameters}},
  year         = 2016,
  journal      = {Journal of the Physical Society of Japan},
  volume       = 85,
  number       = 9,
  pages        = {091014},
  url          = {https://doi.org/10.7566/JPSJ.85.091014}
}

@article{kpoint1976,
  author       = {Monkhorst, Hendrik J. and Pack, James D.},
  title        = {{Special points for Brillouin-zone integrations}},
  year         = 1976,
  month        = {Jun},
  journal      = {Phys. Rev. B},
  publisher    = {American Physical Society},
  volume       = 13,
  pages        = {5188--5192},
  doi          = {10.1103/PhysRevB.13.5188},
  url          = {https://link.aps.org/doi/10.1103/PhysRevB.13.5188}
}

@article{pbe1996,
  author       = {Perdew, John P. and Burke, Kieron and Ernzerhof, Matthias},
  title        = {{Generalized Gradient Approximation Made Simple}},
  year         = 1996,
  month        = {Oct},
  journal      = {Phys. Rev. Lett.},
  publisher    = {American Physical Society},
  volume       = 77,
  pages        = {3865--3868},
  doi          = {10.1103/PhysRevLett.77.3865},
  url          = {https://link.aps.org/doi/10.1103/PhysRevLett.77.3865}
}

@article{qe2017,
  author       = {Giannozzi, P and Andreussi, O and Brumme, T and Bunau, O and Buongiorno Nardelli, M and Calandra, M and Car, R and Cavazzoni, C and Ceresoli, D and Cococcioni, M and Colonna, N and Carnimeo, I and Dal Corso, A and de Gironcoli, S and Delugas, P and DiStasio, R A and Ferretti, A and Floris, A and Fratesi, G and Fugallo, G and Gebauer, R and Gerstmann, U and Giustino, F and Gorni, T and Jia, J and Kawamura, M and Ko, H-Y and Kokalj, A and Küçükbenli, E and Lazzeri, M and Marsili, M and Marzari, N and Mauri, F and Nguyen, N L and Nguyen, H-V and Otero-de-la-Roza, A and Paulatto, L and Poncé, S and Rocca, D and Sabatini, R and Santra, B and Schlipf, M and Seitsonen, A P and Smogunov, A and Timrov, I and Thonhauser, T and Umari, P and Vast, N and Wu, X and Baroni, S},
  title        = {{Advanced capabilities for materials modelling with Quantum ESPRESSO}},
  year         = 2017,
  month        = {oct},
  journal      = {Journal of Physics: Condensed Matter},
  publisher    = {IOP Publishing},
  volume       = 29,
  number       = 46,
  pages        = 465901,
  doi          = {10.1088/1361-648X/aa8f79},
  url          = {https://dx.doi.org/10.1088/1361-648X/aa8f79}
}

@article{PhysRevB.41.7892,
  author       = {Vanderbilt, David},
  title        = {{Soft self-consistent pseudopotentials in a generalized eigenvalue formalism}},
  year         = 1990,
  month        = {Apr},
  journal      = {Phys. Rev. B},
  publisher    = {American Physical Society},
  volume       = 41,
  pages        = {7892--7895},
  doi          = {10.1103/PhysRevB.41.7892},
  url          = {https://link.aps.org/doi/10.1103/PhysRevB.41.7892},
  issue        = 11,
  numpages     = {0}
}

@article{Momma:ko5060,
  author       = {Momma, Koichi and Izumi, Fujio},
  title        = {{{\it VESTA}: a three-dimensional visualization system for electronic and structural analysis}},
  year         = 2008,
  month        = {Jun},
  journal      = {Journal of Applied Crystallography},
  volume       = 41,
  number       = 3,
  pages        = {653--658},
  doi          = {10.1107/S0021889808012016},
  url          = {https://doi.org/10.1107/S0021889808012016}
}

@article{Izumi2007ThreeDimensionalVI,
  author       = {Fujio Izumi and K Momma},
  title        = {{Three-Dimensional Visualization in Powder Diffraction}},
  year         = 2007,
  journal      = {Solid State Phenomena},
  volume       = 130,
  pages        = {15--20},
  url          = {https://api.semanticscholar.org/CorpusID:136681366}
}

@article{PhysRevB.84.144416,
  author       = {Steele, Andrew J. and Baker, Peter J. and Lancaster, Tom and Pratt, Francis L. and Franke, Isabel and Ghannadzadeh, Saman and Goddard, Paul A. and Hayes, William and Prabhakaran, D. and Blundell, Stephen J.},
  title        = {{Low-moment magnetism in the double perovskites Ba${}_{2}$$M$OsO${}_{6}$ ($M=\text{Li},\text{Na}$)}},
  year         = 2011,
  month        = {Oct},
  journal      = {Phys. Rev. B},
  publisher    = {American Physical Society},
  volume       = 84,
  pages        = 144416,
  doi          = {10.1103/PhysRevB.84.144416},
  url          = {https://link.aps.org/doi/10.1103/PhysRevB.84.144416},
  issue        = 14,
  numpages     = 4
}

@article{PhysRevLett.108.177202,
  author       = {Garc\'{\i}a-Flores, A. F. and Moreira, A. F. L. and Kaneko, U. F. and Ardito, F. M. and Terashita, H. and Orlando, M. T. D. and Gopalakrishnan, J. and Ramesha, K. and Granado, E.},
  title        = {{Spin-Electron-Phonon Excitation in Re-based Half-Metallic Double Perovskites}},
  year         = 2012,
  month        = {Apr},
  journal      = {Phys. Rev. Lett.},
  publisher    = {American Physical Society},
  volume       = 108,
  pages        = 177202,
  doi          = {10.1103/PhysRevLett.108.177202},
  url          = {https://link.aps.org/doi/10.1103/PhysRevLett.108.177202},
  issue        = 17,
  numpages     = 5
}

@article{PhysRevLett.98.017204,
  author       = {Azimonte, C. and Cezar, J. C. and Granado, E. and Huang, Q. and Lynn, J. W. and Campoy, J. C. P. and Gopalakrishnan, J. and Ramesha, K.},
  title        = {{Incipient Orbital Order in Half-Metallic ${\mathrm{Ba}}_{2}{\mathrm{FeReO}}_{6}$}},
  year         = 2007,
  month        = {Jan},
  journal      = {Phys. Rev. Lett.},
  publisher    = {American Physical Society},
  volume       = 98,
  pages        = {017204},
  doi          = {10.1103/PhysRevLett.98.017204},
  url          = {https://link.aps.org/doi/10.1103/PhysRevLett.98.017204},
  issue        = 1,
  numpages     = 4
}

@article{PhysRevB.59.11159,
  author       = {Kobayashi, K.-I. and Kimura, T. and Tomioka, Y. and Sawada, H. and Terakura, K. and Tokura, Y.},
  title        = {{Intergrain tunneling magnetoresistance in polycrystals of the ordered double perovskite ${\mathrm{Sr}}_{2}{\mathrm{FeReO}}_{6}$}},
  year         = 1999,
  month        = {May},
  journal      = {Phys. Rev. B},
  publisher    = {American Physical Society},
  volume       = 59,
  pages        = {11159--11162},
  doi          = {10.1103/PhysRevB.59.11159},
  url          = {https://link.aps.org/doi/10.1103/PhysRevB.59.11159},
  issue        = 17,
  numpages     = {0}
}

@article{PhysRevB.64.125126,
  author       = {Wu, Hua},
  title        = {{Electronic structure study of double perovskites ${A}_{2}{\mathrm{FeReO}}_{6}(A=\mathrm{B}\mathrm{a},\mathrm{S}\mathrm{r},\mathrm{C}\mathrm{a})$ and ${\mathrm{Sr}}_{2}M{\mathrm{MoO}}_{6}$ $(M=\mathrm{C}\mathrm{r},\mathrm{M}\mathrm{n},\mathrm{F}\mathrm{e},\mathrm{C}\mathrm{o})$ by LSDA and $\mathrm{LSDA}+U$}},
  year         = 2001,
  month        = {Sep},
  journal      = {Phys. Rev. B},
  publisher    = {American Physical Society},
  volume       = 64,
  pages        = 125126,
  doi          = {10.1103/PhysRevB.64.125126},
  url          = {https://link.aps.org/doi/10.1103/PhysRevB.64.125126},
  issue        = 12,
  numpages     = 7
}

@article{PhysRevMaterials.7.084409,
  author       = {Cong, Rong and Garcia, Erick and Forino, Paola C. and Tassetti, Anna and Allodi, Giuseppe and Reyes, Arneil P. and Tran, Phoung M. and Woodward, Patrick M. and Franchini, Cesare and Sanna, Samuele and Mitrovi\ifmmode \acute{c}\else \'{c}\fi{}, Vesna F.},
  title        = {{Effects of charge doping on Mott insulator with strong spin-orbit coupling, ${\mathrm{Ba}}_{2}{\mathrm{Na}}_{1\ensuremath{-}x}{\mathrm{Ca}}_{x}{\mathrm{OsO}}_{6}$}},
  year         = 2023,
  month        = {Aug},
  journal      = {Phys. Rev. Mater.},
  publisher    = {American Physical Society},
  volume       = 7,
  pages        = {084409},
  doi          = {10.1103/PhysRevMaterials.7.084409},
  url          = {https://link.aps.org/doi/10.1103/PhysRevMaterials.7.084409},
  issue        = 8,
  numpages     = 16
}

@article{PhysRevB.102.195424,
  author       = {Basini, M. and Sanna, S. and Orlando, T. and Bordonali, L. and Cobianchi, M. and Arosio, P. and Mariani, M. and Peddis, D. and Bonanni, V. and Mathieu, R. and Kalaivani, T. and Singh, G. and Larionova, J. and Guari, Y. and Lartigue, L. and Lascialfari, A.},
  title        = {{Low-temperature anomalies in muon spin relaxation of solid and hollow $\ensuremath{\gamma}\text{\ensuremath{-}}\mathrm{F}{\mathrm{e}}_{2}{\mathrm{O}}_{3}$ nanoparticles: A pathway to detect unusual local spin dynamics}},
  year         = 2020,
  month        = {Nov},
  journal      = {Phys. Rev. B},
  publisher    = {American Physical Society},
  volume       = 102,
  pages        = 195424,
  doi          = {10.1103/PhysRevB.102.195424},
  url          = {https://link.aps.org/doi/10.1103/PhysRevB.102.195424},
  issue        = 19,
  numpages     = 9
}

@article{NISHIBORI20011045,
  author       = {E. Nishibori and M. Takata and K. Kato and M. Sakata and Y. Kubota and S. Aoyagi and Y. Kuroiwa and M. Yamakata and N. Ikeda},
  title        = {{The large Debye–Scherrer camera installed at SPring-8 BL02B2 for charge density studies}},
  year         = 2001,
  journal      = {Nuclear Instruments and Methods in Physics Research Section A: Accelerators, Spectrometers, Detectors and Associated Equipment},
  volume       = {467-468},
  pages        = {1045--1048},
  doi          = {https://doi.org/10.1016/S0168-9002(01)00639-8},
  issn         = {0168-9002},
  url          = {https://www.sciencedirect.com/science/article/pii/S0168900201006398},
  note         = {Proceedings of the 7th Int. Conf. on Synchrotron Radiation Instru mentation}
}

@article{10.1063/1.4999454,
  author       = {Kawaguchi, S. and Takemoto, M. and Osaka, K. and Nishibori, E. and Moriyoshi, C. and Kubota, Y. and Kuroiwa, Y. and Sugimoto, K.},
  title        = {{High-throughput powder diffraction measurement system consisting of multiple MYTHEN detectors at beamline BL02B2 of SPring-8}},
  year         = 2017,
  month        = {08},
  journal      = {Review of Scientific Instruments},
  volume       = 88,
  number       = 8,
  pages        = {085111},
  doi          = {10.1063/1.4999454},
  issn         = {0034-6748},
  url          = {https://doi.org/10.1063/1.4999454},
}

@article{PhysRevB.95.064416,
  author       = {Ahn, Kyo-Hoon and Pajskr, Karel and Lee, Kwan-Woo and Kune\ifmmode \check{s}\else \v{s}\fi{}, Jan},
  title        = {{Calculated $g$-factors of $5d$ double perovskites ${\mathrm{Ba}}_{2}{\mathrm{NaOsO}}_{6}$ and ${\mathrm{Ba}}_{2}{\mathrm{YOsO}}_{6}$}},
  year         = 2017,
  month        = {Feb},
  journal      = {Phys. Rev. B},
  publisher    = {American Physical Society},
  volume       = 95,
  pages        = {064416},
  doi          = {10.1103/PhysRevB.95.064416},
  url          = {https://link.aps.org/doi/10.1103/PhysRevB.95.064416},
  issue        = 6,
  numpages     = 5
}

@article{PhysRevB.65.144413,
  author       = {Wiebe, C. R. and Greedan, J. E. and Luke, G. M. and Gardner, J. S.},
  title        = {{Spin-glass behavior in the $S=1/2$ fcc ordered perovskite ${\mathrm{Sr}}_{2}{\mathrm{CaReO}}_{6}$}},
  year         = 2002,
  month        = {Mar},
  journal      = {Phys. Rev. B},
  publisher    = {American Physical Society},
  volume       = 65,
  pages        = 144413,
  doi          = {10.1103/PhysRevB.65.144413},
  url          = {https://link.aps.org/doi/10.1103/PhysRevB.65.144413},
  issue        = 14,
  numpages     = 9
}

@article{Barbosa2022,
  author       = {da Cruz Pinha Barbosa, Victor and Xiong, Jie and Tran, Phuong Minh and McGuire, Michael A. and Yan, Jiaqiang and Warren, Matthew T. and Aguilar, Rolando Valdes and Zhang, Wenjuan and Randeria, Mohit and Trivedi, Nandini and Haskel, Daniel and Woodward, Patrick M.},
  title        = {{The Impact of Structural Distortions on the Magnetism of Double Perovskites Containing $5d^1$ Transition-Metal Ions}},
  year         = 2022,
  journal      = {Chemistry of Materials},
  volume       = 34,
  number       = 3,
  pages        = {1098--1109},
  doi          = {10.1021/acs.chemmater.1c03456},
  url          = {https://doi.org/10.1021/acs.chemmater.1c03456},
}

@article{PhysRevB.104.024437,
  author       = {Svoboda, Christopher and Zhang, Wenjuan and Randeria, Mohit and Trivedi, Nandini},
  title        = {{Orbital order drives magnetic order in $5{d}^{1}$ and $5{d}^{2}$ double perovskite Mott insulators}},
  year         = 2021,
  month        = {Jul},
  journal      = {Phys. Rev. B},
  publisher    = {American Physical Society},
  volume       = 104,
  pages        = {024437},
  doi          = {10.1103/PhysRevB.104.024437},
  url          = {https://link.aps.org/doi/10.1103/PhysRevB.104.024437},
  issue        = 2,
  numpages     = 12
}

@article{GARRITY2014446,
  author       = {Kevin F. Garrity and Joseph W. Bennett and Karin M. Rabe and David Vanderbilt},
  title        = {{Pseudopotentials for high-throughput DFT calculations}},
  year         = 2014,
  journal      = {Computational Materials Science},
  volume       = 81,
  pages        = {446--452},
  doi          = {https://doi.org/10.1016/j.commatsci.2013.08.053},
  issn         = {0927-0256},
  url          = {https://www.sciencedirect.com/science/article/pii/S0927025613005077}
}

@article{PhysRevB.110.064425,
  author       = {Onuorah, Ifeanyi John and Frassineti, Jonathan and Wang, Qiaochu and Isah, Muhammad Maikudi and Bonf\`a, Pietro and Rau, Jeffrey G. and Rodriguez-Rivera, J. A. and Kolesnikov, A. I. and Mitrovi\ifmmode \acute{c}\else \'{c}\fi{}, Vesna F. and Sanna, Samuele and Plumb, Kemp W.},
  title        = {{Unraveling the magnetic ground state in the alkali-metal lanthanide oxide ${\mathrm{Na}}_{2}\mathrm{Pr}{\mathrm{O}}_{3}$}},
  year         = 2024,
  month        = {Aug},
  journal      = {Phys. Rev. B},
  publisher    = {American Physical Society},
  volume       = 110,
  pages        = {064425},
  doi          = {10.1103/PhysRevB.110.064425},
  url          = {https://link.aps.org/doi/10.1103/PhysRevB.110.064425},
  issue        = 6,
  numpages     = 12
}

@article{Pourovskii2016,
  author       = {Pourovskii, L V.},
  title        = {{Two-site fluctuations and multipolar intersite exchange interactions in strongly correlated systems}},
  year         = 2016,
  month        = {Sep},
  journal      = {Phys. Rev. B},
  publisher    = {American Physical Society},
  volume       = 94,
  pages        = 115117,
  doi          = {10.1103/PhysRevB.94.115117},
  url          = {https://link.aps.org/doi/10.1103/PhysRevB.94.115117},
  issue        = 11,
  numpages     = 13
}

@article{Fioremosca2024b,
  author       = {Fiore Mosca, Dario and Franchini, Cesare and Pourovskii, Leonid V.},
  title        = {{Interplay of superexchange and vibronic effects in the hidden order of {Ba$_2$MgReO$_6$} from first principles}},
  year         = 2024,
  month        = {Nov},
  journal      = {Phys. Rev. B},
  publisher    = {American Physical Society},
  volume       = 110,
  pages        = {L201101},
  doi          = {10.1103/PhysRevB.110.L201101},
  url          = {https://link.aps.org/doi/10.1103/PhysRevB.110.L201101},
  issue        = 20,
  numpages     = 7
}

@article{hubbard_1,
  author       = {J. Hubbard},
  title        = {{Electron correlations in narrow energy bands}},
  year         = 1963,
  journal      = {Proc. Roy. Soc. (London)},
  volume       = {A 276},
  pages        = 238
}

@article{Georges1996,
  author       = {Georges, A. and Kotliar, G. and Krauth, W. and Rozenberg, M. J.},
  title        = {{Dynamical mean-field theory of strongly correlated fermion systems and the limit of infinite dimensions}},
  year         = 1996,
  journal      = {Rev. Mod. Phys.},
  publisher    = {American Physical Society},
  volume       = 68,
  pages        = {13--125}
}

@article{Anisimov1997_1,
  author       = {V I Anisimov and A I Poteryaev and M A Korotin and A O Anokhin and G Kotliar},
  title        = {{First-principles calculations of the electronic structure and spectra of strongly correlated systems: dynamical mean-field theory}},
  year         = 1997,
  journal      = {Journal of Physics: Condensed Matter},
  volume       = 9,
  pages        = 7359,
  nonum        = 35
}

@article{Lichtenstein_LDApp,
  author       = {Lichtenstein, A. I. and Katsnelson, M. I.},
  title        = {{Ab initio calculations of quasiparticle band structure in correlated systems: {LDA++} approach}},
  year         = 1998,
  journal      = {Phys. Rev. B},
  volume       = 57,
  pages        = {6884--6895},
  issue        = 12,
  numpages     = {0}
}

@article{Aichhorn2016,
  author       = {Aichhorn, M. and Pourovskii, L V. and Seth, P. and Vildosola, V. and Zingl, M. and Peil, O. E and Deng, X. and Mravlje, J. and Kraberger, G. J and Martins, C. and others},
  title        = {{TRIQS/DFTTools: A TRIQS application for ab initio calculations of correlated materials}},
  year         = 2016,
  journal      = {Computer Physics Communications},
  publisher    = {Elsevier},
  volume       = 204,
  pages        = {200--208}
}

@book{Wien2k,
  author       = {P Blaha and K Schwarz and G Madsen and D Kvasnicka and J Luitz and R Laskowski and  F Tran and Marks, L D.},
  title        = {{WIEN2k, An augmented Plane Wave + Local Orbitals Program for Calculating Crystal Properties}},
  year         = 2018,
  publisher    = {Karlheinz Schwarz, Techn. Universität Wien, Austria,ISBN 3-9501031-1-2}
}

@article{Pourovskii2021,
  author       = {Pourovskii, L V. and Fiore Mosca, D. and Franchini, C.},
  title        = {{Ferro-octupolar Order and Low-Energy Excitations in ${\mathrm{d}}^{2}$ Double Perovskites of Osmium}},
  year         = 2021,
  month        = {Nov},
  journal      = {Phys. Rev. Lett.},
  publisher    = {American Physical Society},
  volume       = 127,
  pages        = 237201,
  doi          = {10.1103/PhysRevLett.127.237201},
  url          = {https://link.aps.org/doi/10.1103/PhysRevLett.127.237201},
  issue        = 23,
  numpages     = 6
}

@misc{magint,
  author       = {Pourovskii, L. V. and  Fiore Mosca, D.},
  title        = {{MagInt}},
  howpublished = {\url{https://github.com/MagInteract/MagInt}}
}

@article{fioremosca2025,
  title = {Antiferro octupolar order in the $5{d}^{1}$ double perovskite {Sr$_2$MgReO$_6$} and its spectroscopic signatures},
  author = {Fiore Mosca, Dario and Pourovskii, Leonid V.},
  journal = {Phys. Rev. Res.},
  volume = {7},
  issue = {3},
  pages = {L032016},
  numpages = {6},
  year = {2025},
  month = {Jul},
  publisher = {American Physical Society},
  doi = {10.1103/tvp5-mpy9},
  url = {https://link.aps.org/doi/10.1103/tvp5-mpy9}
}

@article{Rotter2004,
  author       = {M Rotter},
  title        = {{Using McPhase to calculate magnetic phase diagrams of rare earth compounds}},
  year         = 2004,
  journal      = {Journal of Magnetism and Magnetic Materials},
  volume       = {272-276},
  pages        = {E481-E482},
  doi          = {https://doi.org/10.1016/j.jmmm.2003.12.1394},
  issn         = {0304-8853},
  url          = {https://www.sciencedirect.com/science/article/pii/S0304885303022881},
  note         = {Proceedings of the International Conference on Magnetism (ICM 2003)}
}

@article{PhysRevLett.133.066501,
  author       = {Agrestini, S. and Borgatti, F. and Florio, P. and Frassineti, J. and Fiore Mosca, D. and Faure, Q. and Detlefs, B. and Sahle, C. J. and Francoual, S. and Choi, J. and Garcia-Fernandez, M. and Zhou, K.-J. and Mitrovi\ifmmode \acute{c}\else \'{c}\fi{}, V. F. and Woodward, P. M. and Ghiringhelli, G. and Franchini, C. and Boscherini, F. and Sanna, S. and Moretti Sala, M.},
  title        = {{Origin of Magnetism in a Supposedly Nonmagnetic Osmium Oxide}},
  year         = 2024,
  month        = {Aug},
  journal      = {Phys. Rev. Lett.},
  publisher    = {American Physical Society},
  volume       = 133,
  pages        = {066501},
  doi          = {10.1103/PhysRevLett.133.066501},
  url          = {https://link.aps.org/doi/10.1103/PhysRevLett.133.066501},
  issue        = 6,
  numpages     = 7
}

@article{Kugel_Khomskii_1982,
  author       = {Kliment I Kugel' and D I Khomskiĭ},
  title        = {{The Jahn-Teller effect and magnetism: transition metal compounds}},
  year         = 1982,
  month        = {apr},
  journal      = {Soviet Physics Uspekhi},
  publisher    = {},
  volume       = 25,
  number       = 4,
  pages        = 231,
  doi          = {10.1070/PU1982v025n04ABEH004537},
  url          = {https://dx.doi.org/10.1070/PU1982v025n04ABEH004537},
}

@article{Pelka2013,
  author       = {R. Pe{\l}ka and P. Konieczny and M. Fitta and M. Czapla and P. M. Zieli{\'n}ski and M. Ba{\l}anda and T. Wasiuty{\'n}ski and Y. Miyazaki and A. Inaba and D. Pinkowicz and B. Sieklucka},
  title        = {{Magnetic systems at criticality: Different signatures of scaling}},
  year         = 2013,
  journal      = {Acta Physica Polonica A},
  volume       = 124,
  number       = 5,
  pages        = {977--983},
  doi          = {10.12693/APhysPolA.124.977}
}

@article{PhysRevB.82.012407,
  author       = {Baker, P. J. and Lancaster, T. and Franke, I. and Hayes, W. and Blundell, S. J. and Pratt, F. L. and Jain, P. and Wang, Z.-M. and Kurmoo, M.},
  title        = {{Muon spin relaxation investigation of magnetic ordering in the hybrid organic-inorganic perovskites $[{({\text{CH}}_{3})}_{2}{\text{NH}}_{2}]M{(\text{HCOO})}_{3}$ $(M=\text{Ni},\text{Co},\text{Mn},\text{Cu})$}},
  year         = 2010,
  month        = {Jul},
  journal      = {Phys. Rev. B},
  publisher    = {American Physical Society},
  volume       = 82,
  pages        = {012407},
  doi          = {10.1103/PhysRevB.82.012407},
  url          = {https://link.aps.org/doi/10.1103/PhysRevB.82.012407},
  issue        = 1,
  numpages     = 4
}

@article{Pourovskii2025,
  author       = {Pourovskii, Leonid V. and Fiore Mosca, Dario and Celiberti, Lorenzo and Khmelevskyi, Sergii and Paramekanti, Arun and Franchini, Cesare},
  title        = {{Hidden orders in spin--orbit-entangled correlated insulators}},
  year         = 2025,
  month        = {Sep},
  day          = {01},
  journal      = {Nature Reviews Materials},
  volume       = 10,
  number       = 9,
  pages        = {674--696},
  doi          = {10.1038/s41578-025-00824-z},
  issn         = {2058-8437},
  url          = {https://doi.org/10.1038/s41578-025-00824-z}
}

@FOOTNOTE{Note1,key="Note1",note="See Supplemental Material at [URL will be inserted by publisher], which includes Refs.~\cite {Bonfa2018_muesr, Pourovskii2016, magint, Fioremosca2025, Fioremosca2024b, PhysRevB.103.104401}, for more additional data and details"}
\end{document}